%% file: paper.tex
\documentclass[letterpaper,twocolumn,10pt]{article}

\usepackage{usenix}
\usepackage[l2tabu, orthodox]{nag}
\usepackage{hyperref} 
\usepackage{cite}

\input{macros.tex}

\newcommand{\ie}{\textit{i.e.}\xspace}
\newcommand{\eg}{\textit{e.g.}\xspace}
\newcommand{\ea}{\textit{et al.}\xspace}

\begin{document}

% Don't want date printed
\date{}

\title{
    {\Large \textbf{How Do Tor Users Interact With Onion Services?}}
}

\author{
	\textup{\phantom{ZZZZ}Philipp Winter\phantom{ZZZZ}} \\
	Princeton University
	\and
	\textup{\phantom{ZZZZ}Anne Edmundson\phantom{ZZZZ}} \\
	Princeton University
	\and
	\textup{\phantom{ZZZ}Laura M. Roberts\phantom{ZZZ}} \\
	Princeton University
	\and
	\textup{Agnieszka Dutkowska-Żuk} \\
	Independent
	\and
	\textup{\phantom{ZZZZ}Marshini Chetty\phantom{ZZZZ}} \\
	Princeton University
	\and
	\textup{\phantom{ZZZZ}Nick Feamster\phantom{ZZZZ}} \\
	Princeton University
}

\maketitle

\thispagestyle{empty}

\input{abstract}

\input{introduction}

\input{background}
\input{related-work}
\input{method}
\input{results}

\input{discussion}

\input{conclusion}

\input{acknowledgments}

\balance
\bibliographystyle{abbrv}
{\footnotesize
\bibliography{paper-bibtex}
}

\end{document}

%% file: macros.tex
\usepackage{subcaption}
\usepackage[nameinlink]{cleveref}
\usepackage{tikz}
\usepackage[T1]{fontenc}
\usepackage[scaled=0.8]{beramono}
\usepackage[scaled=0.8]{berasans}
\usepackage{enumitem} % For custom symbols in enumerations.
\usepackage{wasysym}  % For \Square and \Circle.
\usepackage[english]{babel}
\usepackage{booktabs}
\usepackage{enumitem} % For custom symbols in enumerations.
\usepackage{wasysym}  % For \Square and \Circle.
\usepackage[utf8]{inputenc}
\usepackage{csquotes}
\usepackage{xspace}
\usepackage{balance}
\usepackage[font=small,format=plain,labelfont=bf,textfont=it,aboveskip=4pt,belowskip=-12pt]{caption}
\usepackage{microtype}
\usepackage{url}

\graphicspath{ {./figures/} }

\usetikzlibrary{positioning}

\usepackage{xcolor} % For colorful links.

\usepackage[nameinlink]{cleveref}
\usepackage{fontawesome} % For the link icon in the bibliography.

\newcommand{\first}{(\textit{i})}
\newcommand{\second}{(\textit{ii})}
\newcommand{\third}{(\textit{iii})}

\newcommand{\mc}[1]{\textcolor{blue}{\noindent[MC: #1]}}
\newcommand{\adz}[1]{\textcolor{magenta}{\noindent[ADZ: #1]}}
\newcommand{\lmr}[1]{\textcolor{magenta}{\noindent[LMR: #1]}}
\newcommand{\annie}[1]{\textcolor{magenta}{\noindent[AE: #1]}}
\newcommand{\nf}[1]{\textcolor{magenta}{\noindent[NF: #1]}}

\renewcommand{\mc}[1]{\textcolor{blue}{}}
\renewcommand{\lmr}[1]{\textcolor{magenta}{}}
\renewcommand{\adz}[1]{\textcolor{magenta}{}}
\renewcommand{\annie}[1]{\textcolor{magenta}{}}
\renewcommand{\nf}[1]{\textcolor{magenta}{}}

\renewcommand{\paragraph}[1]{\vspace*{0.03in}\noindent{\bf #1}}

\definecolor{darkblue}{rgb}{0.1,0.1,0.4}
\hypersetup{
	pdftitle={How Do Tor Users Interact with Onion Services?},
	pdfauthor={Philipp Winter, Anne Edmundson, Laura M. Roberts, Marshini Chetty, and Nick Feamster},
	pdfkeywords={Tor, usability, anonymity, usable security},
	colorlinks=true,
	urlcolor=darkblue,
	linkcolor=darkblue,
	citecolor=darkblue
}

%% file: abstract.tex
\subsection*{Abstract}
Onion services are anonymous network services that are exposed over the Tor
network.  In contrast to conventional Internet services, onion services are
private, generally not indexed by search engines, and use self-certifying
domain names that are long and difficult for humans to read. In this paper, we
study how people perceive, understand, and use onion services based on
data from 17 semi-structured interviews and an online survey of 517 users. We
find that users have an incomplete mental model of onion services, use these services for anonymity 
and have varying trust in onion services in general. Users also have difficulty 
discovering and tracking onion sites and authenticating them. Finally, users want
technical
improvements to onion services and better information
on how to use them. Our findings suggest various
improvements for the security and usability of Tor onion services,
including ways to automatically detect phishing of onion services, more clear security
indicators, and ways to manage onion domain names that are difficult to remember.

%% file: introduction.tex
\section{Introduction}
\label{sec:introduction}

The Tor Project's onion services provide a popular way of
running an anonymous network service. 
In contrast to anonymity for clients (\eg, obfuscating
a client IP address using a virtual private network), Tor onion services provide
anonymity for servers, allowing a web server to obfuscate its network location 
(specifically, its IP address). An operator of a web service may need to anonymize
the location of a web service to escape harassment, speak out against power, or
voice dissenting opinions.

Onion services were originally developed in 2004 and have  recently seen
growing numbers of both servers and users.  As of June 2018, The Tor
Project's statistics count more than 100,000 onion services each day, collectively
serving traffic at a rate of nearly 1~Gbps.  In addition to web sites, onion services
include
metadata-free instant messaging~\cite{ricochet} and file
sharing~\cite{onionshare}.  The Tor Project currently does not have data on
the number of onion service users, but Facebook reported in 2016 that more
than one million users logged into its onion service in one month~\cite
{facebook-users}.

Onion services differ from conventional web services in four ways;
First, they can only be accessed over the Tor network. Second, onion domains are
hashes over their public key, which make them difficult to remember. Third, the
network
path between client and the
onion service is typically longer, increasing latency and thus reducing the performance
of the service. Finally, onion services are private by default, meaning that
users must discover these sites organically, rather than with a search engine.

In this paper, we study how users cope with these idiosyncrasies, by exploring the
following questions:
\begin{itemize}
	\itemsep=-1pt
\item What are users' mental models of onion services? 
\item How do users use and manage onion
services? 
\item What are the challenges of using onion services?
\end{itemize}
\noindent
Because onion services depend on the Tor Browser and the underlying Tor network
to exchange traffic, some of our study also explored users' mental models of Tor
itself, but this topic is not the focus of our paper.

To answer these questions, we employed a mixed-methods approach. First, we conducted exploratory
interviews with Tor and onion service users to guide the design of an online
survey. We then conducted a large-scale online survey that included
questions on Tor Browser, onion service usage and operation, onion site
phishing, and users' general expectations of privacy. Next, we conducted
follow-up interviews to further explore the topics and themes that we discovered
in the exploratory
interviews and survey. We complemented this qualitative data with an analysis of
``leaked''
DNS lookups to onion domains, as seen from a DNS root server; this data gave us
insights into actual usage patterns and allowed us to corroborate some of the findings
from the interviews and surveys.

We find that many Tor users misunderstand technical aspects
of onion services, such as the nature of the domain format, rendering these
users more vulnerable to phishing attacks. Second, we find that users have
many issues using and managing onion services, including having trouble discovering
and tracking new onion domains.
Our data also
suggests
that users may visit onion domains that are slight variations of popular onion domains,
suggesting that typos or phishing attacks may occur on onion domains.
Third, users want improvements to onion services such as improved performance and
easier ways to keep track of and verify onion domains as authentic. 
Many of the shortcomings that we discover could be addressed with straightforward
and immediate improvements to
the Tor Browser, including improved security indicators and mechanisms to automatically
detect domains that may be typos or phishing attacks.

Tor is currently testing the next generation of
onion services, which will address various security issues and upgrade to faster,
future-proof cryptography.  The findings from our work can inform the design of
privacy and security enhancements to onion services and Tor Browser at a critical
time as these improvements are being deployed. 
This paper makes the following contributions:
\begin{itemize}
\itemsep=-2pt 
\item  We provide new, large-scale empirical evidence from Tor
users that sheds light on how these users perceive, use, and manage
onion services. Our work confirms and extends previous findings
on Tor Browser users' mental models~\cite{Forte2017a}.
 \item We provide empirical evidence that characterizes onion domain name
lookups based on a dataset from the {\tt .onion} requests from DNS B root, both
extending previous work on onion domain usage~\cite{thomas2014measuring, mohaisen2017leakage}
and corroborating our findings about usability and security problems that we identified
in the survey and interview data.
    \item Based on our findings, we identify usability obstacles to the
        adoption of onion services and suggest possible design enhancements, including
        publishing mechanism for onion services and a Tor Browser extension
        that allows its users to securely and privately bookmark onion domains.
\end{itemize}
\noindent
All code, data, and auxiliary resources are available at \url{https://nymity.ch/onion-services/}.

The rest of this paper is structured as follows. 
\Cref{sec:background} provides background on onion services, and
\Cref{sec:related-work} presents related work.  \Cref{sec:method} presents the methods
for
our interviews, online survey, and DNS data analysis. \Cref{sec:results} presents
results,
\Cref{sec:discussion} discusses the implications of these findings, and
\Cref{sec:conclusion} concludes.

%% file: background.tex
\section{Background: What Are Onion Services?}
\label{sec:background}

Originally called ``hidden services'', onion services were renamed in
2015 to reflect the fact that they provide more than just the ``hiding'' of a
service~\cite{Johnson2015a}---more importantly, they provide end-to-end security and
self-certifying domain names.  Beyond The Tor Project's nomenclature, the ``web''
of onion services is occasionally referred to as the ``Dark Web''. In this
paper, we use only the term onion services.

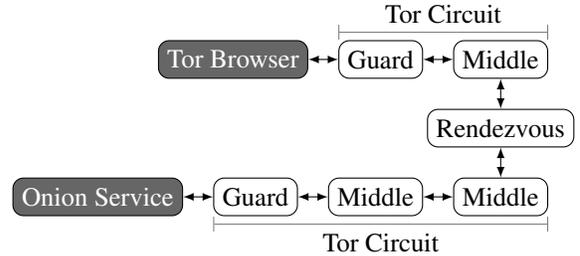
\begin{figure}[t]
\centering
\begin{tikzpicture}[node distance=0.4cm]

\tikzset{>=latex}
\tikzstyle{block} = [rectangle, draw, rounded corners, text centered,
                     minimum height=0.5cm]

\node[block,fill=black!60,
      text=white]         (TB)  {Tor Browser};
\node[block,right=of TB]  (GR1) {Guard};
\node[block,right=of GR1] (MR1) {Middle};
\node[block,below=of MR1] (R)   {Rendezvous};
\node[block,below=of R]   (MR2) {Middle};
\node[block,left=of  MR2] (MR3) {Middle};
\node[block,left=of  MR3] (GR2) {Guard};
\node[block,fill=black!60,
            text=white,
            left=of GR2]  (OS)  {Onion Service};

\draw[<->] (TB.east)   -- (GR1.west);
\draw[<->] (GR1.east)  -- (MR1.west);
\draw[<->] (MR1.south) -- (R.north);
\draw[<->] (R.south)   -- (MR2.north);
\draw[<->] (MR2.west)  -- (MR3.east);
\draw[<->] (MR3.west)  -- (GR2.east);
\draw[<->] (GR2.west)  -- (OS.east);

\draw[|-|,gray] ([yshift=3pt] GR1.north west)
                -- node [above, midway, text=black] {Tor Circuit}
                ([yshift=3pt] MR1.north east);

\draw[|-|,gray] ([yshift=-3pt] GR2.south west)
                -- node [below, midway, text=black] {Tor Circuit}
                ([yshift=-3pt] MR2.south east);

\end{tikzpicture}
\caption{A path to an onion service typically has six Tor relays.
Both the client and the onion service create a Tor circuit (comprising two and
three relays, respectively) to a rendezvous.}
\label{fig:onion-connection}
\end{figure}

Onion services are TCP-based network services that are accessible only over the
Tor network and provide mutual anonymity: the Tor client is anonymous to the server,
and the server is anonymous to the client. 
Clients access onion services via onion domains that are meaningful only 
inside the Tor network.
A path between a client and onion service has six Tor relays by default, as
shown
in \Cref{fig:onion-connection};
the client builds a circuit to a ``rendezvous'' Tor relay,
and the onion service builds a circuit to that same relay.
Neither party learns the other's IP address.

To create an onion domain, a Tor daemon generates an
RSA key pair, computes the SHA-1 hash over the
RSA public key, truncates it to 80 bits, and encodes the result in
a 16-character base32 string (\eg, {\tt expyuzz4wqqyqhjn}).
Because an onion domain is derived directly from its public key, onion domains
are self-certifying: if a client knows a domain, it automatically knows the
corresponding public key.  Unfortunately, this property makes the onion domain difficult
to read, write, or remember. 

As of February 2018, The Tor Project is deploying the next generation of onion
services, whose domains have 56 
characters~\cite[\S~6]{Mathewson2013a} that include a base32 encoding of the
onion service's public key, a checksum, and a version number.   New onion
services will also use elliptic curve cryptography, allowing the entire public
key to be embedded in the domain, as opposed to only the hash of the public
key. These changes will naturally improve the security of onion services but have
important implications for usability, particularly as unreadable onion domain names
get longer.

One way to make onion domains more readable is to repeatedly generate
RSA keys until the resulting domain contains some desired string (\eg, ``facebook'').
These so-called {\em vanity onion domains} include Facebook
({\tt facebookcorewwwi.onion}), ProPublica ({\tt propub3r6espa33w.onion}), and the
New York
Times ({\tt nytimes3xbfgragh.onion}).  
Vanity onion domains still typically have strings of characters that are not meaningful
words, but they may be easier to memorize.
These domains are relatively expensive to create:
given base32's alphabet size of 32
characters, a vanity prefix of length $n$ takes an
average of $0.5 \cdot 32^n$ key creations, 
Given a set of domains that contain a vanity prefix,
one can search this set for a domain that is the easiest to remember, for example
by using a Markov model to filter domains that resemble English words.  The popular
{\tt scallion}
tool~\cite{scallion} parallelizes
the search for vanity domains.

Even if the onion domain is more readable, the user still needs to have a way
of discovering the onion service in the first place. In contrast to
conventional network services, onion services are designed to be difficult to
discover. The operator of an onion service must manually advertise the domain,
for example by manually adding it to onion site search engines such as
Ahmia~\cite{ahmia}.  The lack of a go-to service such as a ``Google for onion
services'' prompted the community to devise various ways to disseminate onion
services through a variety of search engines and curated lists.

\begin{figure}[t]
    \centering

    \begin{subfigure}[t]{\linewidth}
        \centering
        \includegraphics[width=\linewidth]{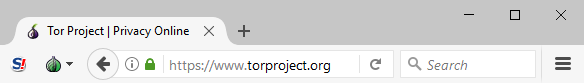}
        \subcaption{Conventional domain.\vspace*{0.15in}}
        \label{fig:non-onion-service}
    \end{subfigure}
    \begin{subfigure}[t]{\linewidth}
        \centering
        \includegraphics[width=\linewidth]{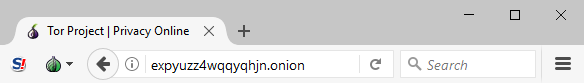}
        \subcaption{Onion service.\vspace*{0.15in}}
        \label{fig:onion-service}
    \end{subfigure}

    \caption{Tor Browser 7.0.10's user interface on Windows 10 when accessing the
    Tor Project website via a
    conventional domain and the corresponding onion service.  
    The onion service lacks a padlock; 
    Tor developers are addressing this issue~\cite{trac23247}.}
\end{figure}

Tor Browser aims to make user access to onion domains seamless.
\Cref{fig:non-onion-service}
shows the interface when accessing The Tor Project's web site;
\Cref{fig:onion-service} shows a connection to the corresponding onion site.
Additionally, because the unreadability of onion domains can make clients 
more susceptible to phishing attacks, website operators who want to provide 
their website as an onion service and do not care about their own 
anonymity can get an extended validation (EV) digital certificate for their 
.onion domain so that clients can be assured that they are connecting to the 
correct site. For example, Facebook’s onion service has a certificate 
associated with it, and this added layer of security is reflected in the Tor Browser.

%% file: related-work.tex
\section{Related Work}
\label{sec:related-work}

%We improve on the existing body of work by studying how users interact with
%onion services specifically.  We believe that some of our findings generalize to other
%systems.  For example, Freenet~\cite{Freenet} and Bitcoin~\cite{Nakamoto2008a}
%employ self-certifying names, too: Freenet uses them in its file naming
%scheme, and Bitcoin uses them in its addressing scheme.

\paragraph{Usage and mental models of Tor Browser.}
Forte \ea studied the privacy practices of contributors to open collaboration
projects such as the Tor Project and Wikipedia to learn about how privacy concerns affect their
contribution practices ~\cite{Forte2017a}. The study, based on 23 interviews, found that contributors worry about an array
of threats, including surveillance, violence, harassment, and loss of
opportunity. This study was not focused on hidden services at all. 
Additionally, Gallagher \ea conducted semi-structured
interviews to understand both why people use Tor Browser and how they understand
the technology~\cite{Gallagher2017a}.  The study found that experts tend to
have a network-centric view of the Tor network and use it frequently,
whereas non-experts have a goal-oriented view and see Tor Browser as a black-box
service. Our work corroborates these findings but is focused on
onion services, rather than generally on Tor Browser.  

\paragraph{Usability of Tor Browser installation.}
Tor Browser has seen many usability improvements since its
creation in 2003~\cite{Syverson2005a}, from a Tor
``button'' to Tor Browser Bundle (now called the Tor Browser).
Ten years ago, Clark \ea used cognitive walkthroughs to
study how users install, configure, and run Tor Browser~\cite{Clark2007a}.
The work revealed hurdles such as jargon-laden documentation,
confusing menus, and insufficient visual feedback.
Norcie \ea\ identified ``stop-points'' in the
installation and use of the Tor Browser Bundle~\cite{Norcie2014a}; these stop-points
require user action but instead cause confusion. the study recommended various changes
to the
installation process and evaluated them in a follow-up study.
Lee \ea~\cite{Lee2017a} studied the usability of Tor Launcher, the graphical
configuration tool that allows users to configure Tor Browser, and found that 79\%
of users' connection attempts in a simulated censored environment failed, but that
various design improvements could reduce these difficulties.

\paragraph{Usability of onion domain names.}
Previous work aimed to improve the usability of onion domain names.
Sai and Fink proposed a mnemonic system that maps 80-bit onion domains to
sentences~\cite{Sai2012a}.  Their work is inspired by mnemonicode,
which maps binary data to words~\cite{mnemonicode}.  Victors \ea designed
the Onion Name System~\cite{Victors2017a}, which
allows users to reference an onion service by a readable, globally unique
identifier.  Kadianakis \ea designed an API that allows
Tor clients to configure name systems (\eg, GNS~\cite{Schanzenbach2012a} or
OnioNS~\cite{Victors2017a}) on a per-domain basis~\cite{Kadianakis2016a}.

\paragraph{Onion domain usage patterns.} If a conventional DNS resolver attempts
to resolve an {\tt .onion} domain (as might happen when a user enters such a domain
name into a normal browser), the resulting DNS lookup for the domain will ``leak''to
the DNS root servers. Previous studies have taken advantage of this leaked information
to characterize the popularity of various onion domains~\cite{mohaisen2017leakage,thomas2014measuring}.
We build on previous work, applying similar analysis with a focus on whether
the lookups suggest usability problems with onion services or the presence of phishing
attacks.

%Forte \ea studied the privacy practices of contributors to open collaboration
%projects~\cite{Forte2017a}.  The authors interviewed 23 contributors to The Tor
%Project and Wikipedia to learn about how privacy concerns affect their
%contribution practices.  The study found that contributors worry about an array
%of threats, including surveillance, violence, harassment, and loss of
%opportunity.

%Inspired by Tor Browser's use as an anti-censorship system, Fifield \ea\ published
%a design to study its usability as a censorship circumvention
%tool~\cite{Fifield2015a}.  Drawing upon both qualitative and quantitative
%methods, the authors plan to recruit hundreds of users to study how they use Tor
%Browser's configuration wizard in an adversarial setting.  

%% file: method.tex
\section{Method}
\label{sec:method}

We used a mixed-methods approach involving interview and survey data, as well as
analysis of DNS query data. 
This section details our interviews (\Cref{sec:interviews}), 
large-scale online survey (\Cref{sec:online-survey}), and the DNS dataset that we
use for our
analysis (\Cref{sec:root-data}).\footnote{Princeton University's institutional review board (IRB) 
approved this study (Protocol \#8251).}

\subsection{Interviews}
\label{sec:interviews}

To help us understand users' mental models of onion services, onion service usage,
and the challenges and benefits of onion services, we conducted qualitative interviews,
which allowed us to design the survey.

\subsubsection{Procedure}

\paragraph{Interview Guide.} We developed a question set that served as the basis
for each
interview,\footnote{The question set is available at
\url{https://nymity.ch/onion-services/pdf/interview-checklist.pdf}.} basing our
design on prior work~\cite{Forte2017a} but focusing particularly
on onion services. The semi-structured nature of our interviews allowed us to deviate
from this question set by asking follow-up questions as appropriate.  

We followed standard consent procedures for
all participants.
We began by asking demographic information (gender, age range, occupation,
country of residence, and level of education), followed by questions about
users' general online behavior. We concluded with questions about Tor Browser
and onion services (\eg, when users started to use these services, how they
track onion links as well as the drawbacks and strengths of these services
based on their own experiences). To gather data about users' mental models of
Tor browser and onion services, we designed a brief sketching exercise similar
to those used in other work~\cite{Poole2008more}. We asked participants to draw
sketches of how they believed Tor and onion services worked and followed up on these
drawings in interviews.

\paragraph{Recruitment.} To select eligible interview subjects, we created a short
pre-interview survey\footnote{The pre-interview survey is available
at \url{https://nymity.ch/onion-services/pdf/pre-interview-survey.pdf}.} asking
users if they were over 18 years of age, if they had used Tor Browser and onion
services, and how they would
rate their general privacy and security knowledge.
To the extent possible, we targeted lay-people and aimed to maximize cultural, gender,
geographic location, education, and age diversity. The Tor Project advertised this
survey both in a blog post~\cite{Winter2017a}
and via Twitter.
We also advertised the study on Princeton's Center for Information Technology (CITP)
blog and
recruited participants in person at an Internet freedom event. 

Recruiting a representative sample of Tor users is difficult, and our recruiting
techniques likely resulted in a biased population for several reasons. First, we
believe that
The Tor Project's blog and Twitter account are followed by disproportionately
more technical users, whereas non-technical users may not generally follow news and updates related to 
Tor via the project's blog and Twitter feed.
Second, Tor users value their privacy more than the
average Internet user, so the users we recruited
may not be as honest and candid about their browsing habits as we would like. 

\paragraph{Interviews.} We conducted 13 interviews in person and four
interviews
remotely---over Skype, Signal, WhatsApp, and Jitsi---depending on the medium that
our participants
preferred.  Two participants
declined to have their interviews
recorded; we recorded the rest of the interviews with the permission of the participant.
All participants answered the interview questions and completed the sketching exercise. Each interview ended with a debriefing
phase to ask if our participants had any remaining questions.  We compensated participants with a \$20 gift card.  We conducted
our first interview on July 13, 2017 and the last on October 20, 2017.  The
median interview time was 34 minutes, with interviews ranging from 20--50 minutes.

\paragraph{Transcription and Analysis.} 
We transcribed our interview recordings
and employed qualitative data coding to analyze the transcripts \cite{seidman2013interviewing}.
In the two cases where we did not have interview recordings, we relied on our field notes.
We developed a codebook based on our research questions and used a combination of
deductive coding 
to identify themes of interest we agreed upon
 and inductive coding to discover emergent phenomena
and to expand the initial codebook. 
We had ten parent codes in total, with examples such as ``Mental model of
onion services'', ``Search habits'', and ``Reasons for using onion services''; and 168
child codes, including ``Definition- anonymous'', ``Word of mouth'', and
``Curiosity''. After we reached consensus on the phenomena of interest, at least
two members of our team (sometimes up to four) read and coded each transcript.
We also held regular research meetings with the entire team of authors to
discuss the coded transcripts and reach consensus on the final themes.

% TODO: Link to themes online.

\subsubsection{Participants}

We interviewed 17 subjects, as summarized in
\Cref{tab:interviewees}.  We only present aggregate demographic information to
protect the identity of our interview participants.  We believe that our sample
is biased towards educated and technical users---almost 60\% of our participants
have a postgraduate degree---but our sample also shows the diversity among Tor's
user
base: our participants comprised human rights activists, legal professionals,
writers, artists, and journalists, among others.
In remainder of the paper, we use the denotation `P'
to refer  to interview participants.

\begin{table*}[t]
	\centering
	\begin{tabular}{l r r | l r r | l r r | l r r}
	\toprule
	{\bf Age} & \# & \% &
	{\bf Gender} & \# & \% &
	{\bf Continent of residence} & \# & \% &
	{\bf Education} & \# & \% \\
	\midrule
	18--25 & 2  & 11.8 & Female & 5  & 29.4 & Asia          & 3 & 17.6 & No degree    & 1  & 5.9 \\
	26--35 & 10 & 58.8 & Male   & 12 & 70.6 & Australia     & 1 &  5.9 & High school  & 3  & 17.7 \\
	36--45 & 4  & 23.5 &        &    &      & Europe        & 4 & 23.5 & Graduate     & 3  & 17.7 \\
	46--55 & 1  & 5.9  &        &    &      & North America & 8 & 47.1 & Postgraduate & 10 & 58.8 \\
	       &    &      &        &    &      & South America & 1 &  5.9 & & & \\
	\bottomrule
	\end{tabular}
	\caption{The distribution over gender, age, country of residence, and
	education for our 17 interview subjects.  We do not show
	per-person demographic information to protect the identity of our interview
	subjects.}
	\label{tab:interviewees}
\end{table*}

\subsection{Online Survey}
\label{sec:online-survey}

Shortly after we conducted our first batch of interviews, we designed, refined,
and launched an online survey to complement our interview data.\footnote{The
full survey is available at
\url{https://nymity.ch/onion-services/pdf/survey-questions.pdf}.}

\subsubsection{Procedure}

\paragraph{Survey Design.} We created our survey in Qualtrics because an unmodified
Tor Browser could display it correctly.  Unfortunately, Qualtrics requires JavaScript, and 
Tor Browser deactivates if it is set to its highest security setting.  
Several users complained about our reliance on JavaScript in the recruitment
blog post comments~\cite{Winter2017a}. All respondents consented to the survey and
confirmed that they
were at least 18 years old.  Our survey
was only available in English, but we targeted an international audience because
Sawaya \ea showed that cultural differences yield different security
behavior~\cite{Sawaya2017a}, and paying attention to these differences is central
to The Tor Project's global mission.

Most of our survey focused on onion services, but we also included usage
questions about Tor in general because Tor Browser is used to access onion services.
Our survey had of 49 questions, most of which were closed-ended
questions. The first set of questions asked for
basic demographic information such as age, gender, privacy and security knowledge
rating, and education level. Next, the survey asked about Tor usage, such as how
frequently the Tor Browser was used. We also asked about onion services
usage in detail, including questions concerning the usability of onion links, how
users track and manage onion domain links, whether (and why) users had ever set
up or operated an onion site, and whether users were aware of onion site phishing
and impersonation. The last set of questions focused on users' general
expectations of privacy and security when using onion services. 
We incorporated four attention checks to measure a respondent's \emph{degree} of
attention~\cite{Berinsky2014a}. To ensure that participants felt comfortable answering
questions, we did not make questions mandatory. The survey took about 15 minutes
to complete.

\paragraph{Survey Testing.} We used cognitive pretesting (sometimes also called
cognitive interviewing) to
improve the wording of our survey questions~\cite{Collins2003a}.  Pretesting
reveals if respondents understand questions consistently and the way we intended
them to be interpreted. Five pre-testers 
helped us iteratively improve the survey; after pre-testing and revisions,
we launched the survey.

\paragraph{Recruitment.} As with our interviews, we advertised our survey in
a blog post on The
Tor Project's blog~\cite{Winter2017a}, on its corresponding Twitter
account, the CITP blog at Princeton, and on three Reddit subforums.\footnote{
\url{https://reddit.com/r/tor/}, \url{https://reddit.com/r/onions/}
\url{https://reddit.com/r/samplesize/}.}  Unlike our interview participants,
our survey respondents were self-selected.  As with interview recruitment, we expect
this recruitment
strategy biased our sample towards engaged users because casual Tor users are
unlikely to follow The Tor Project's social media accounts.

We did not offer incentives for participation because we wanted respondents to be
able to participate anonymously without providing email addresses. Despite the lack
of incentives, we collected enough
responses.  
Our survey ran from August 16--September 11, 2017 (27~days).

\paragraph{Filtering and Analysis.} Some of the survey
responses were low-quality; people may have rushed their answers,
aborted our survey prematurely, or given deliberately wrong answers. 
To mitigate these effects, we excluded participants who either did
not finish the survey
or who failed more than two out of four attention checks. 
We conducted
a descriptive analysis on the survey data.
We also computed correlation coefficients
between every question pair in the survey, which did not yield significant results.
We thus focus on results from the descriptive analysis. Each percentage is reported
out of the total sample; we denote cases when survey participants chose not to respond
as `No Response'.
Two researchers performed a deductive coding pass on the open-ended survey questions based on
our interview codebook and held meetings to reach consensus on
the final themes discussed. In rest of the paper, we denote survey participants
with `S'. 

\subsubsection{Participants}
 We collected 828 responses, but only 604 (73\%) completed the survey, and 517 (62\%) passed at least two attention
checks.  The rest of the paper focuses on these 517 responses.
\Cref{tab:survey-demo} shows the demographics of our survey.  
As we expected, respondents were {young and educated}: more than 71\%
were younger than 36, and 61\% had at least a
graduate or post-graduate degree.  44\% percent also considered themselves at least
highly knowledgeable in matters of Internet privacy and security. 

\begin{table*}[t]
	\centering
	\begin{footnotesize}
	\begin{tabular}{l r r | l r r | l r r | l r r}
	\toprule
	Gender & \# & \% &
	Age & \# & \% &
	Education & \# & \% &
	Domain knowledge & \# & \% \\
	\midrule
	Male   & 438 & 84.7 & 18--25 & 186 & 35.9 & No degree     &  25 &  4.8 & None
	&   1 &  0.2 \\
	Female &  49 &  9.4 & 26--35 & 180 & 34.8 & High school   & 172 & 33.2 & Mild     &  35 &  6.8 \\
	Other  &  25 &  4.8 & 36--45 &  87 & 16.8 & Graduate      & 214 & 41.4 & Moderate & 178 & 34.4 \\
	No Response    &   5 &  1.0 & 46--55 &  43 &  8.3 & Post graduate & 102 & 19.7 & High     & 227 & 43.9 \\
	       &     &      & 56--65 &  16 &  3.1 & No Response          &   4 &  0.4 & Expert             &  75 & 14.5 \\
	       &     &      & $>$ 65 &   3 &  0.6 &               &     &      & No Response               &   1 &  0.2 \\
	       &     &      & No Response    &   2 &  0.4 &               &     &      &                    &     &      \\
	\bottomrule
	\end{tabular}
	\end{footnotesize}
	\caption{The distribution over gender, age, education, and domain knowledge
	of the survey respondents.  Providing demographic
	information was optional, so we lack data for some respondents.}
	\label{tab:survey-demo}
\end{table*}

\subsection{Domain Name Service (DNS) Queries}
\label{sec:root-data}

We analyzed {\tt .onion} domains leaked via the Domain Name System (DNS) to
better understand onion service usage and look for specific evidence of
usability issues (\eg, onion domains with typographical errors, phishing
attacks). Although onion domains are only resolvable inside the Tor network,
Internet users may attempt to access an onion site using a browser that is not
configured to use Tor, resulting in the DNS query for onion domain ``leaking''
to conventional DNS resolvers---and ultimately to a DNS root server. Because
all onion lookups to a conventional DNS server will result in a cache miss,
all leaked onion lookups will ultimately go to a DNS root server. Thus, DNS
root servers see a good sample of leaked onion domains. Our work builds on a
previous analysis of a similar data set that was conducted several years ago
and which was not focused on onion services specifically like our work
\cite{mohaisen2017leakage,thomas2014measuring}.

We obtained about several days of DNS data from the B root server through the
IMPACT Cyber Trust program~\cite{b_root}.  This data has several hundred {\tt
pcap} files, which contain full packet captures with pseudonymized IP
addresses of all DNS traffic to the B root from September 19,
2017 10:00 UTC to September 21, 2017 23:59 UTC. We analyzed the DNS queries
dataset and present our results alongside our findings from the survey and
interview results. We extracted the QNAME of each DNS query, which yielded
15,471 correctly formatted onion domains that were 16 characters long
(representing an 80-bit hash of the owner's public key) had has any letters of
the alphabet and numbers between 2 and 7. These lookups, of course, may not
always correspond to a real onion site, but they do reflect that some machine
issued a DNS query for that onion domain for some reason.

\subsection{Limitations}
\label{sec:limitations}

%A large and representative sample of Tor users is difficult to obtain.  We could
%have worked with The Tor Project to add a link to our experiment on
%Tor
%Browser's landing page to obtain a larger sample
%of Tor users, but we considered this approach too invasive.  Additionally, users
%who
%rarely restart their Tor Browser or pay no attention to the landing page would
%not have seen the link in any case.  

As we previously mentioned, we asked The Tor Project to disseminate our survey on
its blog and Twitter account, which likely yielded 
the following biases.

\paragraph{Non-response bias.}
People who noticed our call for volunteers but decided against participating may
have valued their privacy too much, falsely believed that their perspective is
irrelevant, lacked time, or had other reasons not to participate.  Nevertheless,
non-respondents may exhibit traits that are fundamentally different from those
who did participate.

\paragraph{Survivor bias.}
Our participants generally were able to tolerate Tor Browser's usability
issues, which is why they are still around to tell their tale.  We likely did
not hear from people who decided that Tor Browser was not for
them and were thus unable to tell us what drove them away.  The danger of
survivor bias lies in optimizing the user experience for the subset of people
whose tolerance for inconvenience is higher than the rest.

\paragraph{Self-selection bias.}
Due to the nature of our online survey, participants could voluntarily select
themselves into our set of respondents.  These respondents may be unusually
engaged, technical, and opinionated.  Indeed, the demographic for our online survey
in
\Cref{sec:online-survey} was young and educated; perhaps Tor
Browser's population is young and educated, as well, but we have no way of knowing.

%% file: results.tex
\section{Results}
\label{sec:results}

We organize the presentation of our findings by topic, including how users
\emph{perceive and use} (\Cref{sec:perception}), \emph{manage} (\Cref{sec:manage}), and
\emph{wish to improve} (\Cref{sec:improve}) onion services.  We interleave the
results from our online survey with our interviews and domain name system data as
appropriate.

\subsection{Perception and Use}
\label{sec:perception}

We first explore how users perceive onion site technology and why they
use onion sites.

\subsubsection{Incomplete mental models of onion services }
 
We asked only our interviewees (not our survey participants) about their mental
models of onion services because it is difficult to collect this type of information from a survey. 
This section thus presents results from the interviews only.

\begin{figure}[t]
  \centering
  \includegraphics[width=0.8\linewidth]{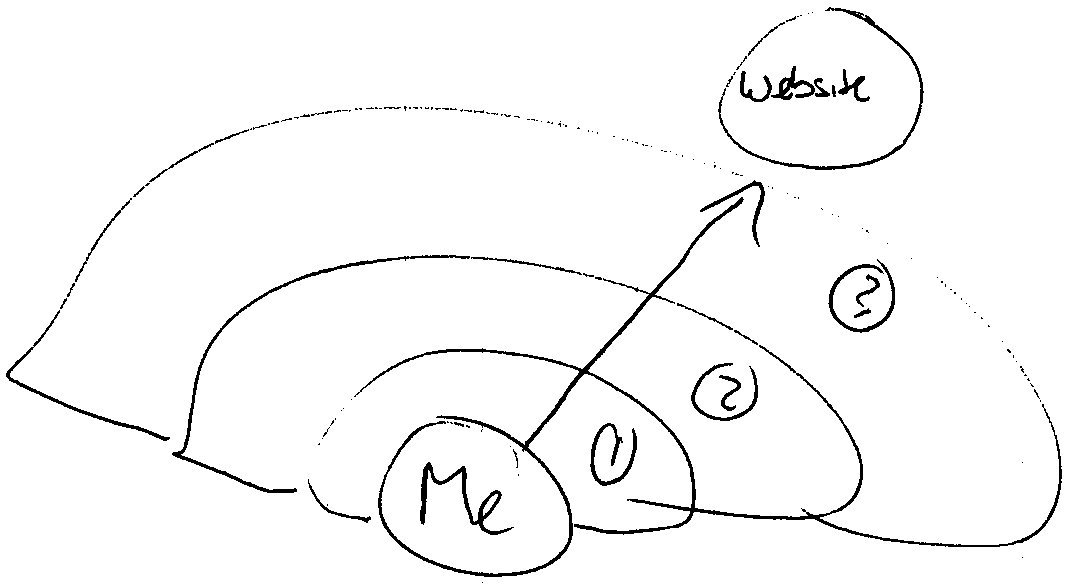}
  \caption{A sketch of interviewee P03's mental model of onion services.  The
  participant referred to several layers of protection.}
  \label{fig:os-sketch}
\end{figure}

\begin{figure}[t]
        \centering
        \includegraphics[width=0.8\linewidth]{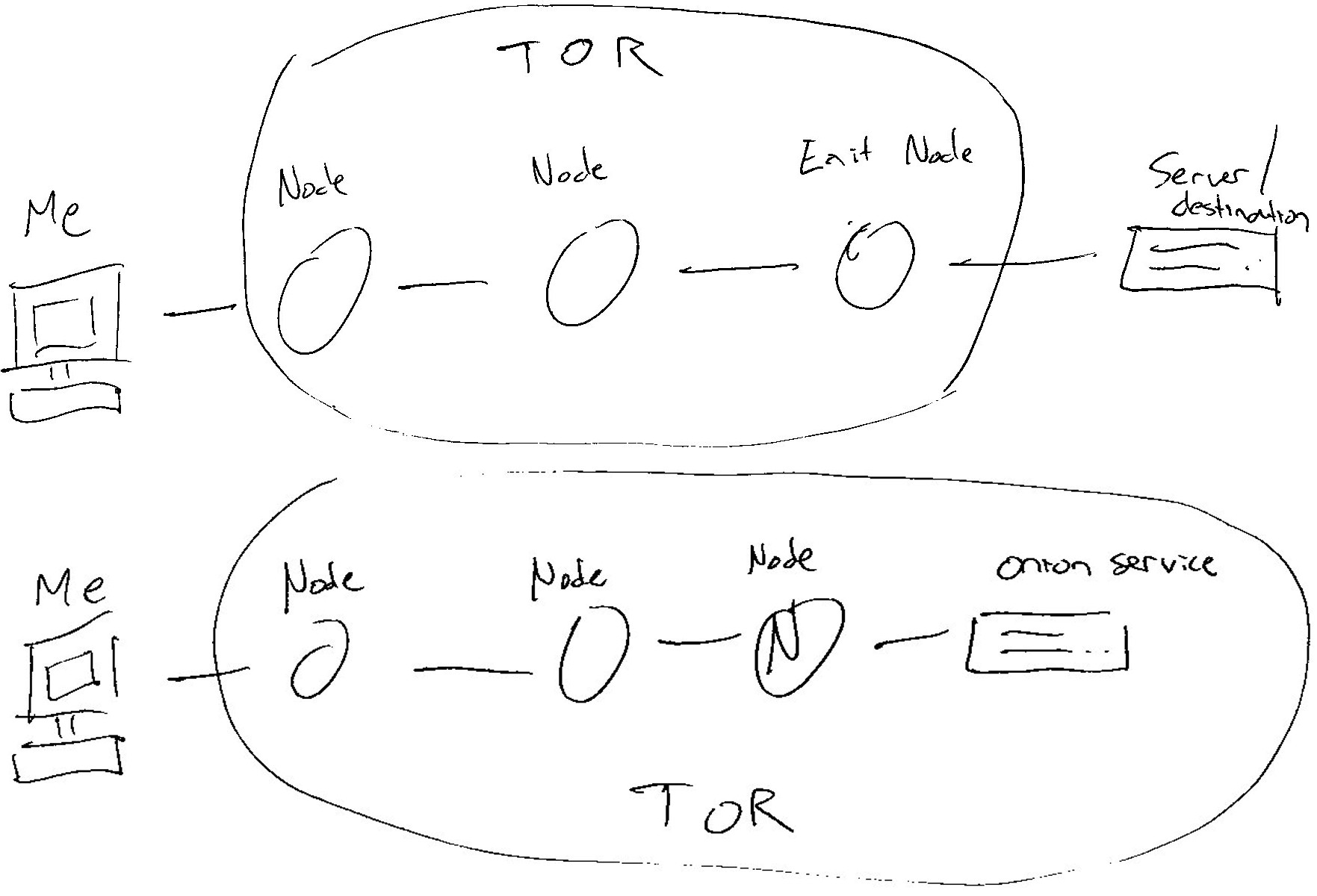}
        \caption{Comparison of two sketches from interviewee P13.  The first
        sketch shows the P13's mental model of Tor and the second one P13's
        mental model of onion services.}
        \label{fig:toros-sketch}
\end{figure}

\paragraph{Perceptions of what an onion service is.} We asked our interview
participants how they defined an onion service, how they work, and what types of
content and services they tend to host.
Terminology was inconsistent and sometimes confusing:
some interviewees referred to onion services
as the dark web and others as hidden services.  (Recall that The Tor Project only
uses the term onion services). 
About half of our interviewees (9/17) knew that onion services enabled a user to
access Web content anonymously. Six interviewees stated that onion
services provide extra layers of protection, an idea that is well-
illustrated in \Cref{fig:os-sketch},\footnote{All sketches are available
online at \url{https://nymity.ch/onion-services/mental-models/}.} and further
elaborated on by participant P03:\emph{``I think it's to do with the different
hops that you build - different layers of making it difficult to find out who
this person is.''} Four interviewees stated that onion services work in a
similar manner to Tor but with different encryption methods, which we can see
on \Cref{fig:toros-sketch}.  A minority of participants had sophisticated
understanding: they referred to the encryption
of data on the end points of a connection; three interviewees
referred to the fact that last hop along the encrypted path
corresponds to an onion link.

\paragraph{Perception of anonymity.} Five interview participants drew the connection between Tor and onion
services, stating that onion services have to be accessed through Tor browser
but at least one did not see any connection between Tor and onion services.
Only three interview participants knew that onion services do not only provide
anonymity to the visitors to a website but also to the onion website  provider
themselves.  In contrast to these interviewees who had some sense of what an
onion service was, nearly half of our interviewees (8/17) were confused about
how to define onion services, were unsure how onion services function or
how to describe them, and did not understand how onion services protect
them. Some of our interviewees did not distinguish disguising their IP address
from disguising their real-world identity and instead used the umbrella term
``anonymity'' to refer to both concepts. This conflation of concepts
paints an incomplete picture of the security and privacy guarantees that the
Tor network provides, with only a few interviewees recognizing that anonymity
is not completely achievable with Tor onion services: \emph{``What's the point of
going to Facebook using onion services when their business model is still
about collecting your data?''} (P7). Other participants simply thought of
onion services as P08 characterized them: \emph{``[the] Internet without
hyperlinks.''} Some of our participants were not aware that
onion services provide end-to-end security and self-certifying names.
Syverson and Boyce explored how onion services can
improve website authentication~\cite{Syverson2015a}, but these benefits are
difficult to convey to non-technical users, and even some experts 
advocated an ``all or nothing'' approach to online anonymity, overlooking
important nuances.

The presence of a large quantity onion domains in the root DNS data \nf{how many?
roughly how many per day? also note that B root is just one slice...}
corroborates prior studies that suggest either Internet
users are attempting to visit an onion domain in a non-Tor browser indicating
a misunderstanding of onion links, that browsers are loading content with
onion links using pre-fetching, or that some web pages or malware are
attempting to load resources from onion sites~\cite{mohaisen2017leakage, thomas2014measuring}.

\paragraph{Perceptions of what an onion service is used for.} Interviewees had various
perceptions of what onion services were used for or why they existed in the first
place. Interviewees sometimes associated onion services with illicit content such as the drug trade or credit card data sales~(2/17) or felt that onion services may be the technology behind anonymous purchases. Similarly, as reported later in the paper, many
survey respondents also voiced concern about illegal and questionable content on onion
services, described by some as a ``Wild West''. Phishing sites, honeypots, and
compromised onion sites further contribute to this perception.

\subsubsection{\mbox{Onion services used mostly for more anonymity}}

\paragraph{Usage.} Our survey asked how often our respondents browse onion
services.  The usage frequency was almost uniformly distributed among our
survey respondents; 24\% use onion sites less than once a month, 22\% use them
about monthly, 25\% weekly, and 23\% daily.  The remaining 6\% had never used
an onion service.  We also asked our interviewees if they had used onion in
the last three months; seven had and seven had not, with four of the latter
group explaining that they had used onion services before, just not in the last
three months.  Only two interviewees had never used onion services
before at all.

\begin{figure}[t]
    \centering
    \includegraphics[width=\linewidth]{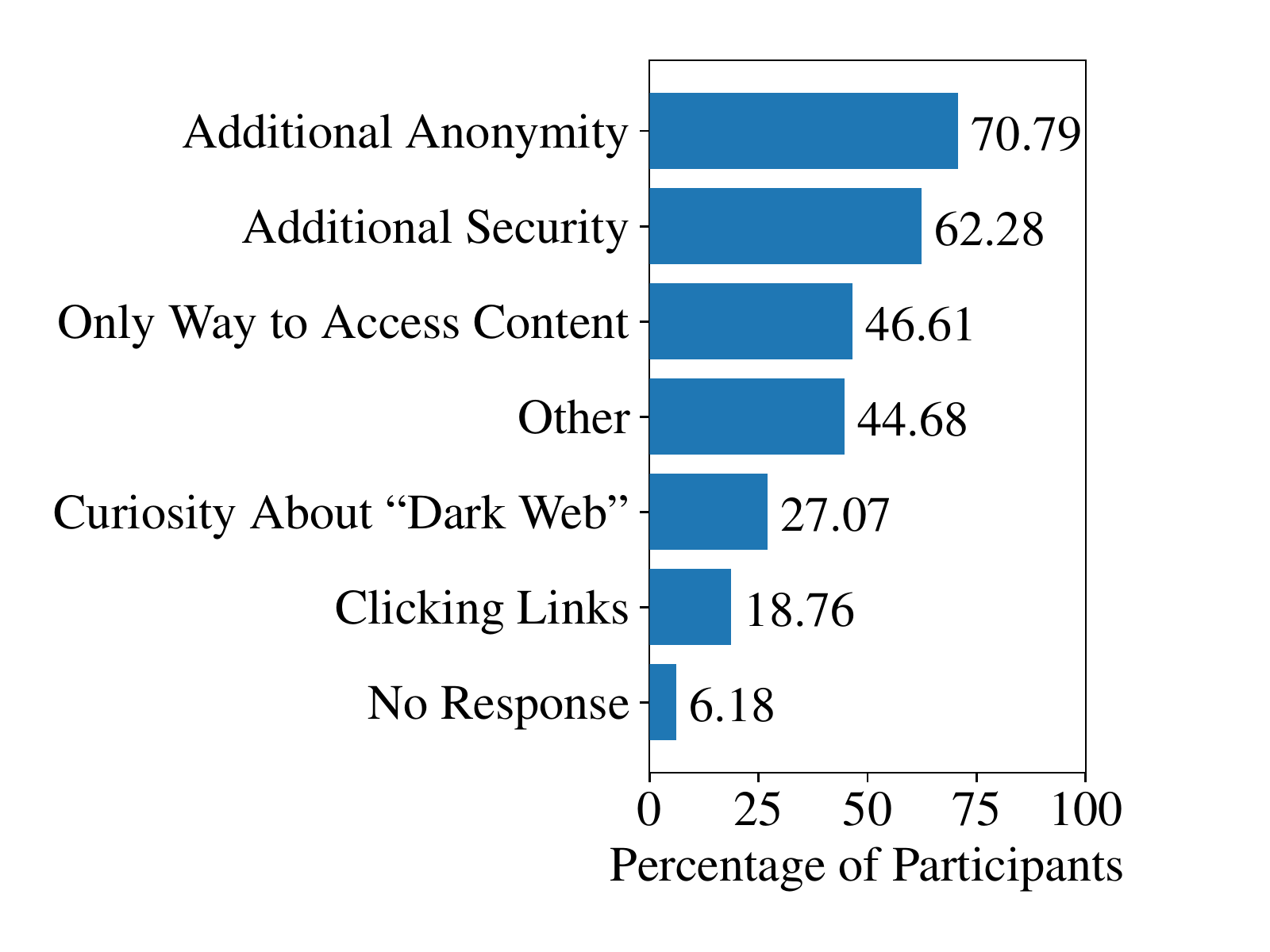}
    \caption{Reasons for using onion services.}
    \label{fig:onion-usage}
\end{figure}

\paragraph{Anonymity and onion service content.} The majority of our survey participants who used onion services did so because of the
additional anonymity (71\%) and the additional security (62\%) (see
\Cref{fig:onion-usage}). For instance, six survey respondents commented on the onion domain format,
indicating that they believed the seemingly-random characters in onion domains
are the reason why onion services are anonymous: \emph{``Onion services stay anonymous through changing their domain, and I
feel that there is a possibility of decreased anonymity with a constant domain
name.'' (S436)}. These participants also believed that vanity domains are ``less anonymous''
because part of their domains is clearly not random.  One survey participant~(S454)
further wrote:\emph{``I understand vanity onion domains are a sign of the
weakness of the hash algorithm used by the Tor network.''} 

Anonymity was also the main reason why our
interviewees used onion services~(6/17).  Another reassuring factor for two of
our interviewees was the feeling of security and safety that onion services
provide.  Furthermore, two interview participants thought of onion services as \emph{``harm
reduction technique.''}  P10 preferred to use Facebook's onion
domain because it impedes tracking efforts. Additionally, 47\% of survey respondents
and three interviewees viewed onion services as the
only way to access content they enjoy, making the use of onion services a
necessity.  

\paragraph{Non-browsing activities.} Of our survey respondents who used onion
services~(485/517), 64\% had these services for purposes
other than web browsing.  Several protocols such as the chat application
Ricochet~\cite{ricochet} and the file sharing application
OnionShare~\cite{onionshare} were purpose-built on top of onion services while
existing TCP-based tools such as {\tt ssh} can transparently use
onion addresses instead of traditional IP addresses. Less than a quarter
 (21\%) of our survey participants used onion services for non-browsing
activities at least once a month such as remote login (ssh) or chat (IRC or XMPP). Our interviewees similarly mentioned using onion services to access
Pirate Bay~(1/17), Ricochet~(1/17), TorChat~(1/17), and OnionShare~(1/17).  

\paragraph{Work or personal reasons.} Survey respondents who selected ``Other''~(45\%)
for onion service usage provided many reasons, including personal~(18/517),
with
the most predominant personal reason being that an onion service gives a machine
behind a network address translation (NAT) device a stable identifier
and can be reached from any other user on the Tor network (there are other ways
to achieve this goal, but for these users, setting up an onion service was the easiest
way). Several interviewees used onion services to accomplish specific tasks.
Five interviewees reported that they use onion services simply for their work,
while four stated personal reasons, such as for a personal blog, or giving someone
access to their home network.  Two interview participants used onion services for
educational purposes.  P3 used onion services to help teach students about the dark web: \emph{``I was teaching a class on Internet technology and
regulations. We were basically showing students how Tor works and part of what
I have to do as a teaching assistant was make students go and basically get to
the moment where they either hire a hitman, buy drugs, or buy weapons.  Just to
show that it's possible. And then obviously we didn't buy it.''}

Other survey respondents reported using onion services to reduce the load on
exit relays, to do technical research, and to access sites that are otherwise
unavailable.  For instance,~7/517 used onion services for hosting a service, one survey respondent admitted using onion services for e-book piracy,  two used onion services as an alternative to a virtual private network and two used them to make their website as private and personal as
they could. 

\paragraph{Exploring the dark web.} 27\% of our survey respondents and two interviewees
wanted to find out
more about the dark web and onion domain content~(3/517) as reasons to use onion services.
Two interviewees used onion services for fun and social reasons---to
\emph{``toy around''}~(P7) and also, as a way of spending time with friends,
as well as to \emph{``show off''} around them by using a technology unfamiliar
to most users. Interestingly, 19\% of survey respondents said that they use onion
services for no particular reason but have clicked on onion links occasionally.

\begin{figure}[t]
    \centering
    \includegraphics[width=\linewidth]{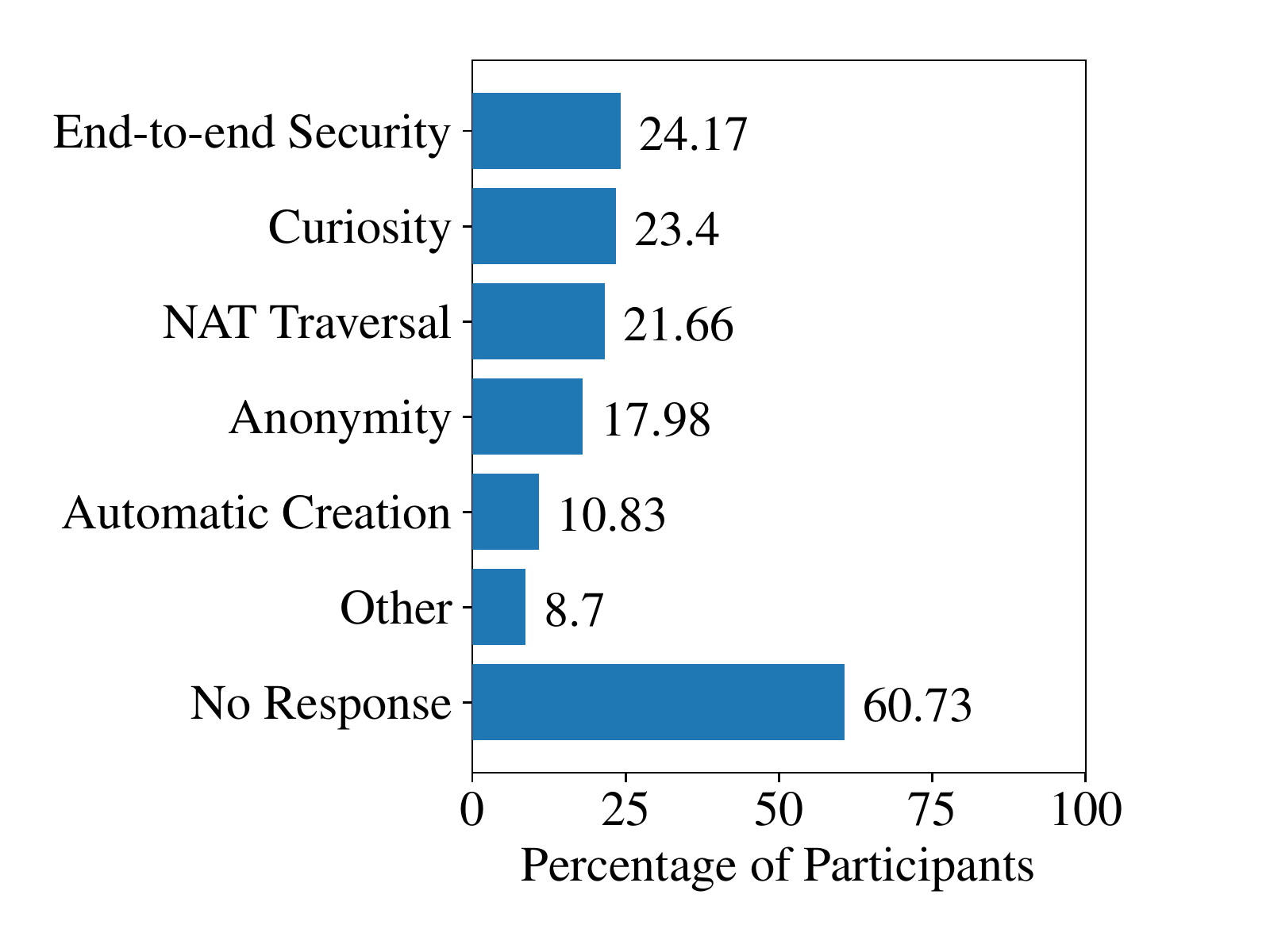}
    \caption{Reasons for running onion services.}
    \label{fig:onion-operation-reasons}
\end{figure}

\subsubsection{\mbox{Onion sites operated for various reasons}}

\paragraph{Setting up an onion service.} 39\% of survey respondents had
set up an onion service at some point. Of the
respondents who had set up onion services of their own (266/517), 31\% had run their onion
service for private use while 21\% had run them for the public. \Cref{fig:onion-operation-reasons} gives an overview of the reasons our respondents
have for running onion services. For instance, the majority of those with
onion services used them for end-to-end security, curiosity, or NAT
traversal. Only 18\% survey respondents had set up onion services for
anonymity, such as to protect their visitors and provide security on their
sites. In the open-ended responses, eleven survey respondents set up onion services
because then their websites
could be accessed from anywhere in the world, and seven survey
respondents set up an onion service simply to test and learn how they work.
Another two survey participants ran onion mirror sites to their personal
websites, and at least one had an onion service as a backup website in case he
lost control over his personal domain. Finally, at least two survey respondents set up
onion for business purposes, work requirements, or to add valuable content to
the onion community.
In a similar vein, at least two interviewees spoke about setting up onion services or using onion services
for work, such as to help Internet users upload leaked documents to their whistleblower website anonymously. In another example, P5 used onion
services in the academic peer review process to allow authors to submit source code or supplementary material anonymously: \emph{``If one of the other reviewers connects to our university site, and we
have some sort of tracking information on there, we would be deanonymizing the
reviewer.  We put it on a Tor hidden service to make sure that the reviewer
remains blind in academic review process.''}

\begin{figure}[t]
    \centering
    \includegraphics[width=\linewidth]{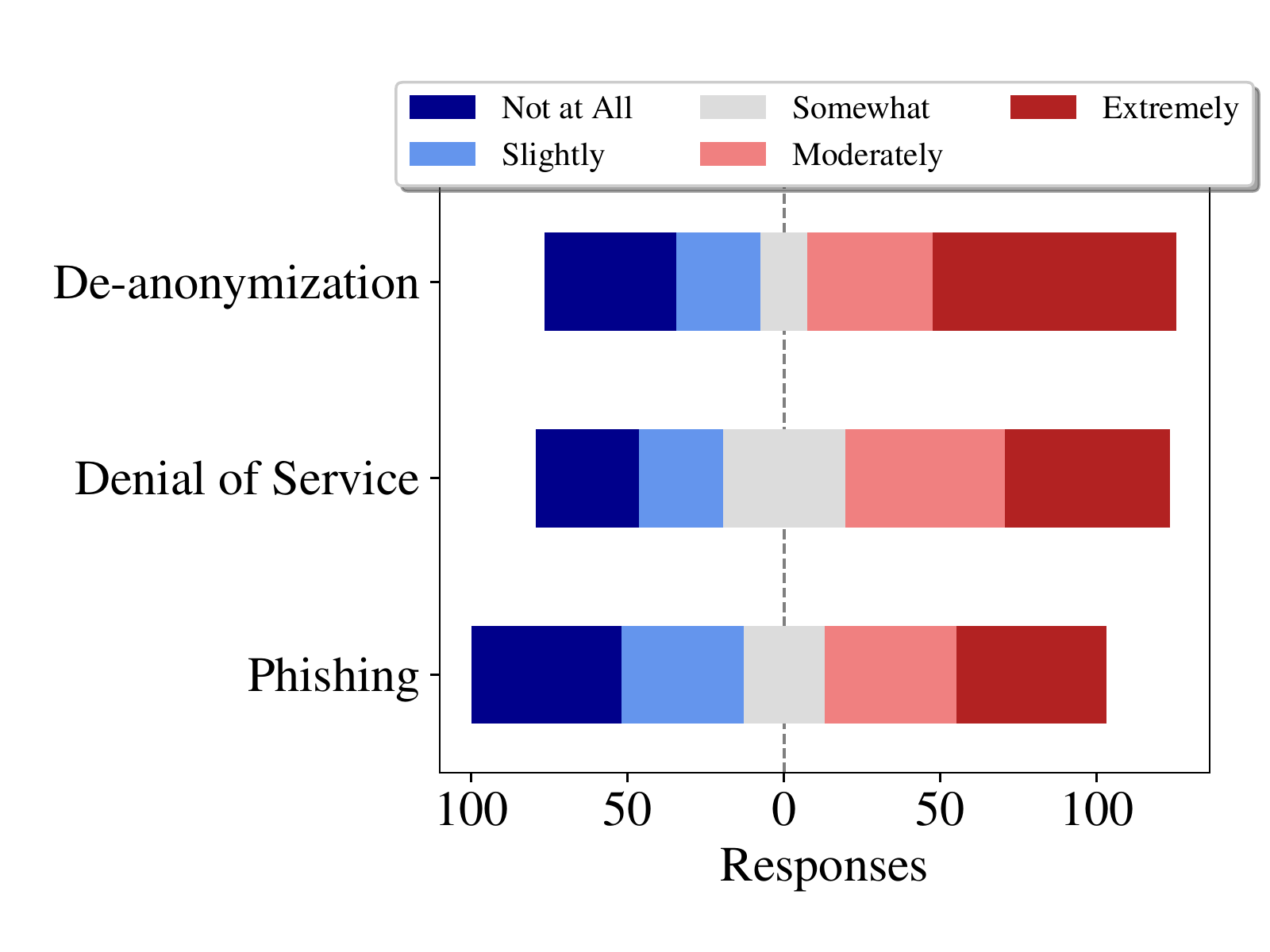}
    \caption{Concerns of onion service operators about attacks.}
    \label{fig:onion-operation-concerns}
\end{figure}

\paragraph{Phishing concerns.} We inquired how concerned the survey
respondents were about three potential attacks on their own onion services:
\first~somebody setting up a phishing site for the operator's site, \second~a
denial-of-service attack, and \third~a deanonymization attack.  According to the
results, shown in \Cref{fig:onion-operation-concerns}, less than 8\% of our
survey respondents who operated an onion service were at least somewhat
concerned about all of these attacks.  Only a small percentage, 15\%, claimed to be extremely
concerned about somebody deanonymizing their onion service, 10\% were extremely concerned about an onion site being taken offline, and only 9\% were concerned about an onion site being impersonated for phishing purposes.  Indeed, in the open-ended responses, we noted that several respondents lamented the difficulty of protecting onion services from
application-layer deanonymization attacks.  Matic \ea\ demonstrated some of
these attacks in 2015~\cite{Matic2015a}.

\subsubsection{Varying trust in Tor and onion services}

\begin{figure}[t]
    \centering
    \includegraphics[width=\linewidth]{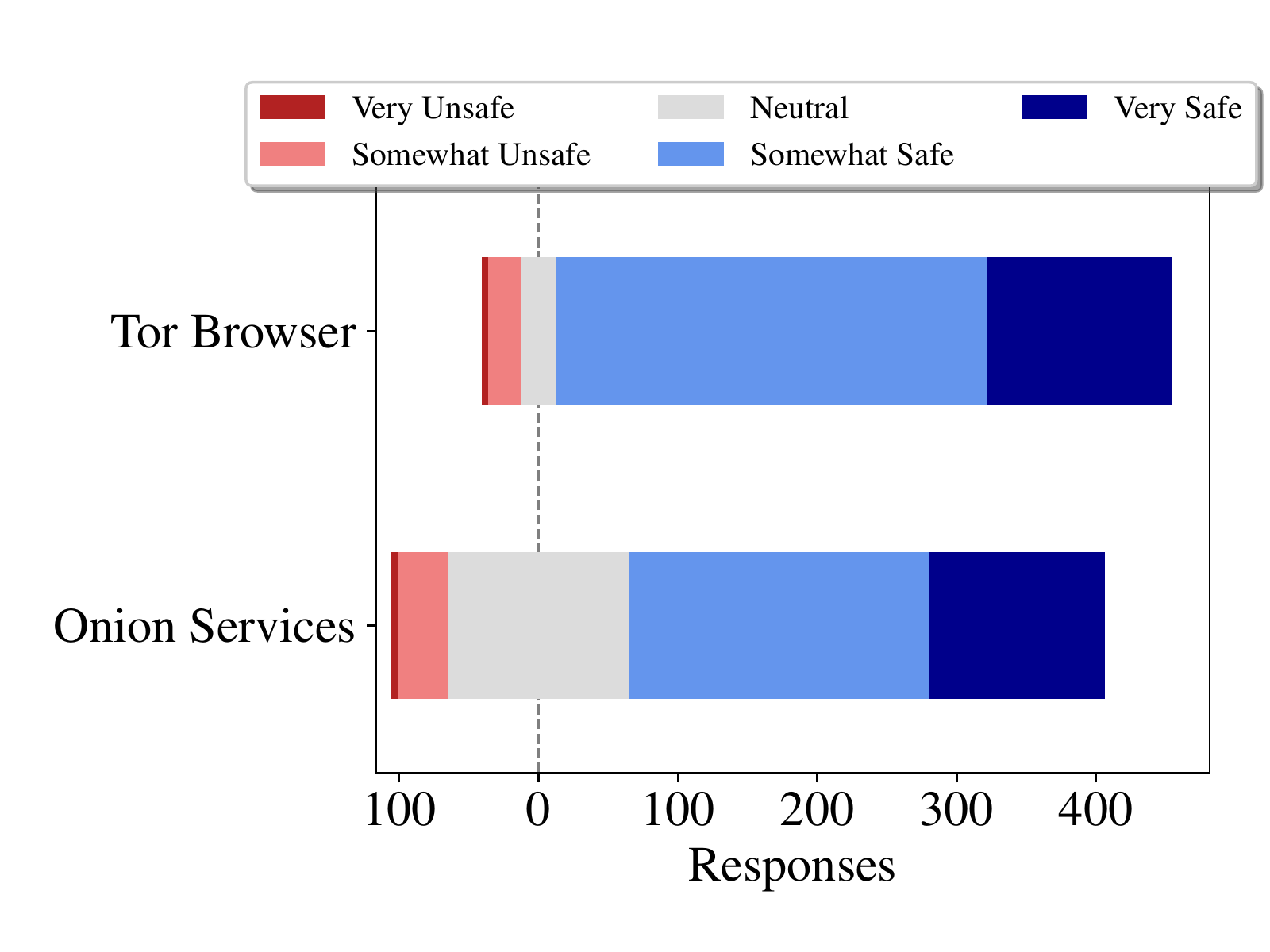}
    \caption{Safety that respondents perceive when using Tor
    Browser and onion services.}
    \label{fig:perceived-security}
\end{figure}

Our survey asked how safe our respondents feel when using Tor Browser and
onion services, respectively.  \Cref{fig:perceived-security} shows that onion services were actually perceived as less safe than Tor browser. 85\% of survey respondents feel at least somewhat safe or very safe using Tor Browser as
compared to only 66\% of onion service users.

\paragraph{Reasons for trust.} Survey responses indicated that participants,
most of whom (85\%) rated themselves as
non-experts (versus 15\% self-rated experts) in knowledge about Internet privacy and security, lacked the ability to evaluate (or even understand)
the Tor network's design which is why they deferred to expert opinion, their gut
feeling, or the trust they place in Tor developers to gauge how much to trust these services. As S450 put it:\emph{`There's a safety tradeoff. My connection to onion sites is more secure from outside eyes, but onion sites are more likely to be scams.'}  With respect to onion
services, the majority of survey respondents expressed that the added security
and anonymity made them feel safe (117/517).  Another factor contributing to the
perceived security of onion services is that advertising companies are nowhere near as present on
onion services as they are on the Web. 80/517 respondents trusted Tor and themselves to be safe on onion services while only a minority of interviewees were content and
believed in the future of onion services~(4/17) or placed their trust in them (2/17).
Additionally, 30/517 participants said they would also choose onion services over regular websites because they trust them.

\paragraph{Reasons for distrust.} 90/517 of survey respondents were skeptical of
trusting onion services because of the possibility of phishing, the fact that onion
services are hard to verify as authentic, and a concern that tracking can still occur even with onion services (59/517). Furthermore, at least 20/517 respondents said their trust of onion services would depend on the content of the services themselves. Some survey respondents did not have a clear understanding of onion services or thought they were the same as regular websites and reported as much~(34/517).

Although our interviewees tended to see onion services as safer than
corresponding websites (eight versus four participants), six participants felt that
users should be careful when using onion services.  Not all participants trusted
onion services~(5/17) and one expressed frustration such as P06:\emph{``I'm pretty 
distrusting with most of the content I access over onion services.
When I want content from a service, I tend to distrust it from the beginning.''}
  Two interviewees
mentioned that websites cannot identify you as the general advantage of onion
services but at least three participants pointed out that websites
actually can determine your identity if you write down your personal details as
well as if you log in into any private accounts while using onion services. Similarly, 20 survey respondents also raised concerned and mentioned not wanting to log in to onion sites because
they believe it defeats the purpose by revealing private data. 

Moreover, one
interview participant (P10) claimed that using onion links may influence the
usability of their ``normal'' corresponding websites---the person shared a story
in which they postulated that their Facebook account had been flagged for
suspicious activity and then was deactivated because they had logged in through
Tor Browser.  These interview participants did not realize that while the
company indeed knows who is logging in, it does not know Tor users' IP
address or operating system.

\subsection{Discovery and Management}
\label{sec:manage}

We now explore how users discover and keep track of onion sites.

\subsubsection{Discovering onion links is not straightforward}
\label{sec:discoveringlinks}

\begin{figure}[t]
    \centering
    \includegraphics[width=\linewidth]{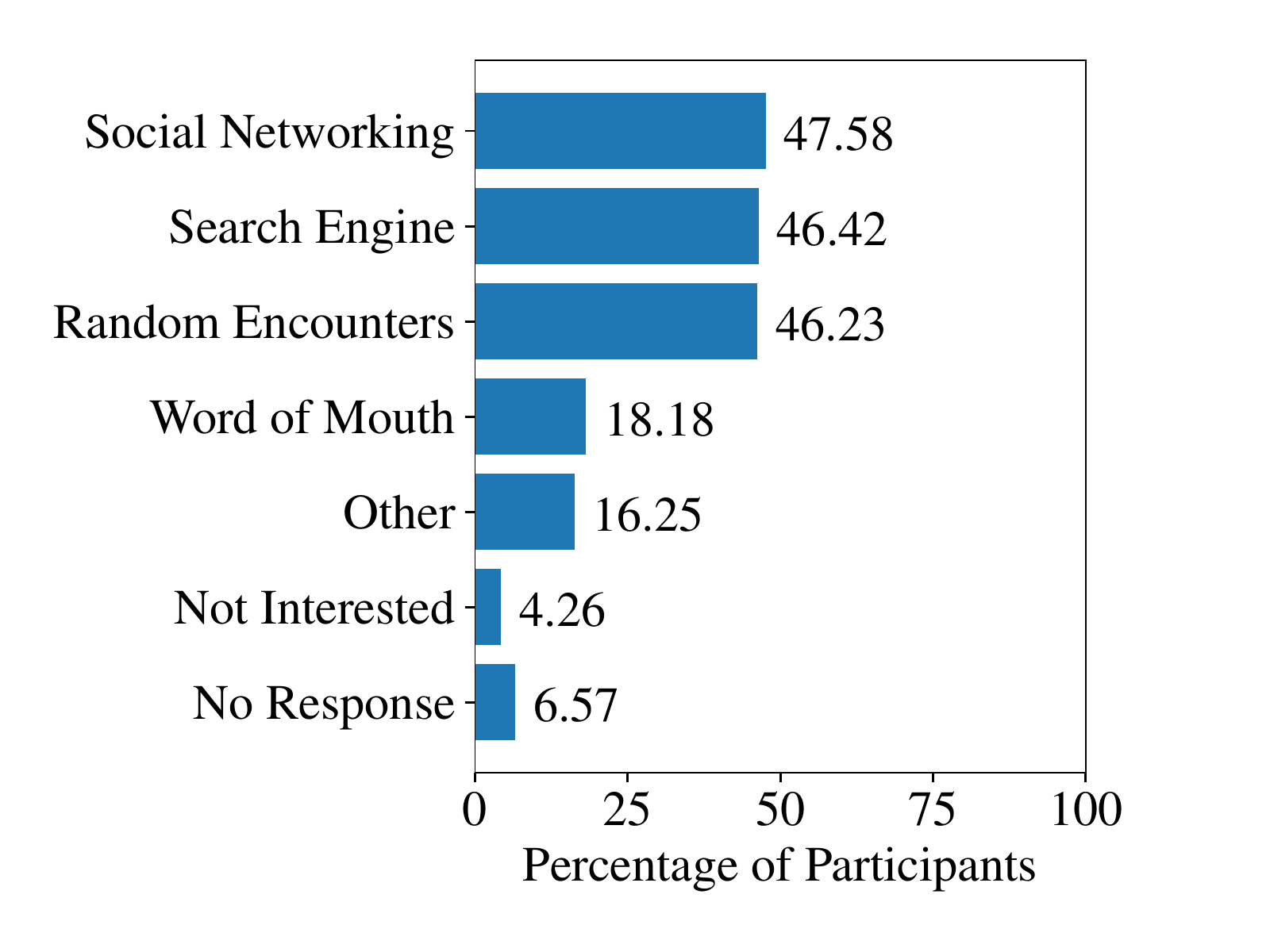}
    \caption{Methods of discovering onion services.}
    \label{fig:onion-discovery}
\end{figure}

Recall that a freshly set up onion service is private by default, leaving it up
to its operator to disseminate the domain.  Established search engines such as
Google are therefore generally inadequate to find content on onion services.  Therefore discovering onion services is not as straightforward as with regular domains  \Cref{fig:onion-discovery} illustrates
the results from our survey.

\paragraph{Social networking site and search engines.} The three most popular ways that almost half of our survey participants
discovered onion sites by were via \first~social
networking sites such as Twitter and Reddit (48\%), \second~search engines such as
Ahmia,\footnote{Ahmia.fi is an onion site search engine that crawls
user-submitted onion domains.  It publishes the list of all indexed onion
services at \url{https://ahmia.fi/onions/}.} (46\%) and \third~randomly encountering
links when browsing the Web (46\%). Survey respondents who selected ``Other'' (16\%) for how they discover onion links predominantly brought
up independently-maintained onion domain aggregators.  A noteworthy example is
the Hidden Wiki used by 13 survey respondents, a community-curated and frequently-forked wiki that contains
categorized links to onion services. At least 34 survey respondents searched for onion links on regular browsers and 18 of these respondents looked specifically at regular websites to see if they had a corresponding onion link. In our interviews, two participants mentioned these techniques too.  Between one to three survey respondents mentioned each of the following: using onion link lists generated by onion spiders, onion.torproject.org, ddg.onion, Imageboard, Google, and even Wikipedia.

We observed similar patterns in our interview
respondents.  Interviewees told us that they find onion links by word of mouth~
(6/17), using a search engine
tool~(5/17) including tools like DuckDuckGo~(1/17), The Pirate Bay~(1/17),
Reddit~(1/17), ahmia.fi~(1/17), and the search widget in the Tor
browser~(1/17).  More of our interviewees discovered onion
services passively (6/17) by just  happening to hear about or know about specific onion
services while five interviewees told us that they looked actively for onion links, browsing for the
content they needed.  

\paragraph{Random encounters or word of mouth.} A significantly less popular discovery
mechanism was discovering links through word of mouth,
which has the advantage that domains come from a trusted source (18\% of survey respondents).   19/517 were frustrated that it was difficult to find out if a regular website had an onion service version even if they visited their website. 
Only 4\% of our
survey respondents---indicated that they were not interested in learning about new
onion services because they only use their own sites (7/517). Similarly, two interviewees
claimed that they never searched for new onion links.

\paragraph{Link discovery challenges.} The majority of our survey respondents (55\%) reported that they were satisfied
with how they discover onion services but a significant proportion of our participants (38\%) were
not and 7\% did not respond to this question. Those satisfied reported that they had no interest in learning about new
onion services, in part because they only use a small set of onion services. Among the survey respondents who were not satisfied with how they discover onion
services (38\%), many (28/517) complained in the open-ended responses about link rot on aggregators where onion links were broken, unusable, or outdated. There is significant churn among onion sites, and our respondents were
frustrated that aggregators are typically not curated and therefore link to
numerous dead domains. The lack of curation also leads to these aggregators'
containing the occasional scam and phishing site.  The difficulty of telling
apart two given onion domain names exacerbates this issue for users. 15/517 did
not trust onion link lists because it is hard to validate if they are legitimate
or not. 28/517 complained about filtering onion sites related
to their interests with several wanting to avoid illegal
and pornographic content, which is often difficult if the description is vague
and the onion domain reveals nothing about its content. For this reason, 5/517 wished aggregators were more verbose in their description
of onion sites.

\paragraph{Lack of good search engines.} Many survey respondents complained about the lack of good search engines (33/517) and
were not aware of search engines such as Ahmia.  Among survey respondents
who were aware of such engines, many were dissatisfied with both the search
results and the number of indexed onion sites.  Unsurprisingly, a \emph{``Google for
onion sites''} was a frequent wish. 
Similarly, one of the biggest issues for our interview participants was that
onion sites are hard to find~(5/17), or as P13 put it: \emph{``How do you find stuff
if you don't know what you're looking for or only have a vague idea?''}  10 survey
respondents desired a better searching solution for onion services even with recognizing that this would be a tradeoff for security so services should have opt-in and opt-out options for discovery. As summarized by one survey respondent: \emph{``Tor is still like the early 1990s Internet where websites were spread by word of mouth and by lists of links. In Tor, people publish lists of onion sites and I pick the ones I'm interested in.  Every Tor search engine is poor and unreliable. Lists of links like Fresh Onions, while useful, often get out of date quickly, since many onion sites are unreliably hosted.   Tor desperately needs a good search engine to find onion sites and ideally some way of identifying what those sites are about before clicking on them, since we lack that info in the URL.''}~(S339)

\begin{figure}[t]
    \centering
    \includegraphics[width=\linewidth]{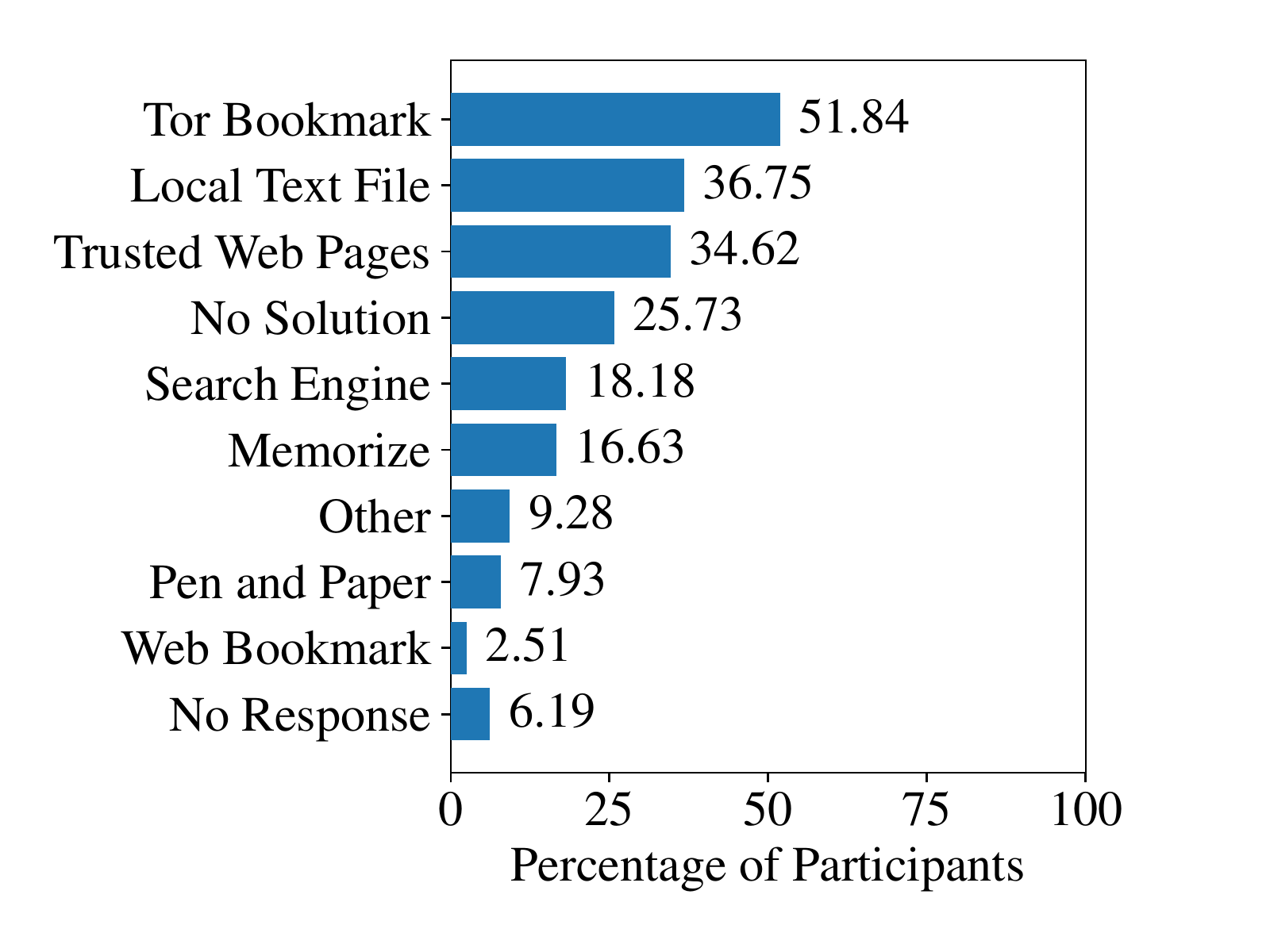}
    \caption{Strategies to manage onion domains.}
    \label{fig:onion-domain-mgmt}
\end{figure}

\subsubsection{Saving and tracking onion links is difficult}

\paragraph{Bookmarking links.} Conventional domains are often easy to remember
and recognize; most onion domains are random strings. We explored how users coped
with this challenge. Most survey respondents (52\%) use
Tor Browser's bookmarks or a web-based bookmarking tool (3\%) to save onion domains as seen in
\Cref{fig:onion-domain-mgmt}. At least two interview participants reported bookmarking links as well.  While
convenient, this method of saving onion links leaves a trace of (presumably)
visited sites on somebody's computer.  One of Tor Browser's security
requirements is ``disk avoidance''---the browser must not write anything to
disk that would reveal the user's browsing history~\cite[\S~2.1]{Perry2017a}.
Bookmarking links is a violation of this security requirement, albeit one that users
seem to want.

\paragraph{Ad-hoc tracking methods.}   Somewhat less popular amongst our survey participants was saving onion
domains in local text files (37\%), getting them from trusted websites (35\%),
using search engines (18\%), memorizing domains (17\%), using some other
techniques (9\%), or employing pen and paper (8\%). Of the 9\% of our survey respondents who selected ``Other'', 15/517 stated that they store onion
domains in an encrypted manner---either in a text file or in their password
manager. Other techniques mentioned by only one or two survey respondents each included using auto-complete, storing them on a personal blog or using Twitter to find links, emailing the links to oneself, using redirect rules to automatically go to the .onion domain, storing the links in a virtual machine, or using Hidden Wiki. Four of our interviewees
reported that they store onion services in a list and three remember (some)
onion services.  Other techniques for saving onion links mentioned by interviewees mirrored those of the survey and included using a Twitter feed to track onion links (1/17) and using
TorChat as storage places for onion links (1/17).  Moreover, one interviewee believed that Tor Browser
remembers onion links and another interview participant (P1) explained: \emph{``The
onion services we run professionally we keep track of because we operate the
server, so that's easy.''}  Notably, just over one-quarter of our survey respondents
(26\%) did
not have a good solution to the problem of tracking onion links and similarly two
interviewees pointed out that they lacked an onion link management mechanism. 

\paragraph{Reaching onion domains quickly.} We also asked our interviewees how they typically reach onion services.  The most
often mentioned technique was copy and pasting domains, done by four
interviewees, followed by three interviewees who simply click on links they
encounter.  Two interviewees would go to onion sites using bookmarks while
another two use Google to get to onion services. Only one interview participant told us that they typed the
domains from their notes. Given the
high number of (possibly insecure) home-baked solutions, a Tor Browser
extension that solves the problem of saving and tracking onion links seems warranted.

\subsubsection{Onion domains are hard to remember}\label{sec:memorize}

\paragraph{Memorization reasons.} Our participants often memorized onion domains to make it easier to visit onion sites and to minimize traces of their browsing habits. Of the survey respondents who memorize onion domains, we found that most
respondents do no memorize any onion domains (60\%) and less than a third (30\%)
memorize one to four onion domains. Only 3\% can memorize more than four domains.
Survey respondents who
memorized domains (65\% of all respondents) did so \first~ automatically because of
typing a domain many times (20\%) \second~to allow them to open an onion site more quickly
(17\%), and \third~to ensure that they are visiting the correct site and not a
phishing site (15\%). Only 9\% were privacy conscious and did so because
bookmarking onion domains leaves a trace. 5\% of the respondents gave other reasons for memorizing onion links. In these open-ended responses, 18 survey participants said that memorizing was simply easy for them, even unintentional. Among these participants, there were only 8/517 that specifically mentioned the Facebook onion site as very easy to remember. Only a few survey respondents  (3/517) did not memorize onion sites at all.

\paragraph{Memorization challenges.} Our interview participants generally found onion domains problematic in terms of having to remember random strings of letters and numbers. Four interviewees perceived onion domains as too long.  Among these participant
was one who further complained about random characters in onion domains.  At least two
interviewees criticized onion links for being hard to remember.  This viewpoint was echoed in our survey, where participants rated URLs such as expyuzz4wqqyqhjn.onion and torproz4wqqyqhjn.onion as
harder to remember because the \emph{``numbers
make the names harder to remember.''} Other survey respondents stated that
vanity domains are easier to remember when they can be pronounced as described in the example quote by survey
respondent (S46): \emph{``phonetic pronunciation
plays a large part in how I remember onions.''} Many other survey
respondents stated that onion domains that are supported by a mnemonic are also
easier to remember; we elaborate on this result in \Cref{sec:vanitymem}.

\subsubsection{\mbox{Vanity domains: more memorable, less trusted}}
\label{sec:vanitymem}

\paragraph{Memorizability.} The majority of our survey respondents appreciated vanity domains because they
were easy to remember (64\%) and easy to recognize (64\%), and they provided a
unique \emph{``branding''} (34\%).  Some survey respondents indicated that a vanity
prefix---like a traditional domain---informs about an onion service's content,
letting visitors know what to expect and thus preventing unpleasant surprises but at least 3/517 wanted more clues to let visitors know more about what the domain content is or for some content to be harder to find. As S423 wrote: \emph{``For less important, high traffic sites (social media like Facebook), it's okay. For sites handling much more sensitive/potentially illicit content, its a good idea to make it difficult to find.''}

\begin{figure}[t]
    \centering
    \includegraphics[width=\linewidth]{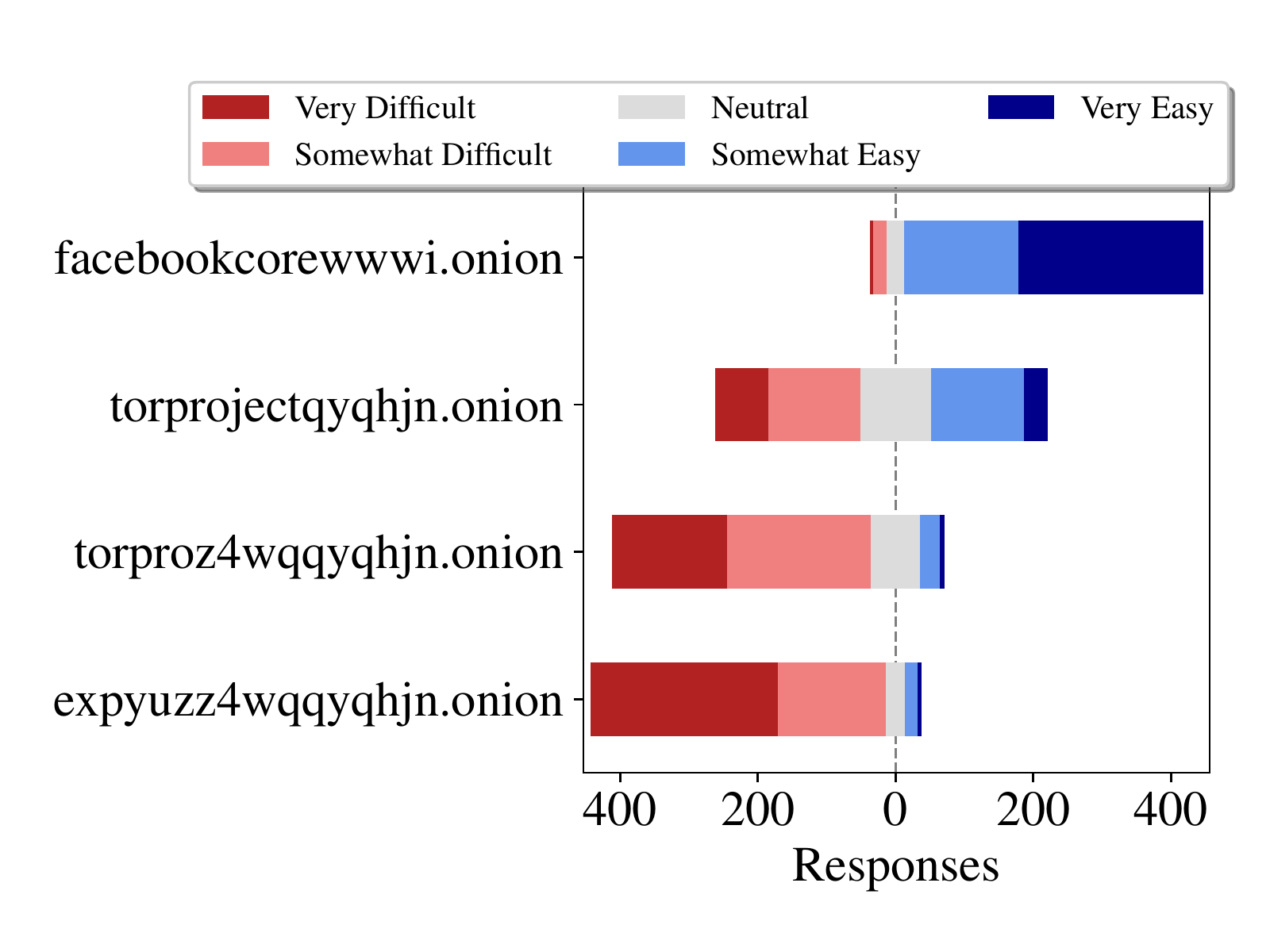}
    \caption{Expected difficulty memorizing four onion domains.}
    \label{fig:memorize-domains}
\end{figure}

Only  15\% did not have an opinion about vanity domains, 8\% reported that they disliked vanity onion domains, and 7\% did not see a benefit of vanity domains.  We 
asked survey respondents about whether or not they memorize vanity
domains---specifically facebookcorewwwi.onion---and how difficult they find it
to memorize onion domains of differing levels of vanity.  Only 20\% of respondents
replied that facebookcorewwwi.onion is among the sites that they have memorized.
This is because it is ``easy to memorize'' (S391) and ``after seeing [it] many times, I
automatically start to memorize it.''(S94) Depending on the format of the vanity
domain, our survey respondents expressed differing levels of ease for memorizing
them; these results are shown in \Cref{fig:memorize-domains}. Most participants found it easier to memorize vanity domains with a longer recognizable prefix such as Facebook's. Interestingly, only 4/517  survey respondents considered vanity
domains economically unfair because wealthy entities can afford to generate
longer prefixes such as Facebook.

\paragraph{Usable links.} Ten out of seventeen interviewees saw vanity domains as a significant usability
improvement to the regular onion domains: \emph{``In terms of mnemonics and easier recollection if you can chunk
words that are associated with daily life and not just a random. If there's
entropy in the stream, there's no way I'm going to remember more than a few
characters''} (P18). P10 had a different perspective that suggested these vanity domains make onion services more usable: \emph{``I think that for people
who don't spend a lot of time using those types of services, it definitely gives
you a more familiar framework for thinking about where you are on the Internet.
If people think \ldots people have a pretty strange geographic metaphors for
navigating the Internet, but I think this idea of where are you?  Well, I'm at
this place I can't even name, I can't say it out loud, I think that can be a
barrier for people.''} 

\paragraph{Phishing and security.} 
If users focus on the vanity part of a domain only, attackers can create an
similar domain that features the original's prefix but differs in
subsequent characters.  Nurmi~\cite{Nurmi2015a} and
Monteiro~\cite{Monteiro2016a} have both documented such an attack, but its
effectiveness is not known.

Indeed, in several cases, both survey (29/517) and interview participants found
that vanity domains were not practical and seemed to distrust them because they felt they made phishing easier: \emph{``I don't
think it's useful because \ldots it's followed by another random word \ldots and
phishing can still copy that \ldots I don't think what I can remember is safe
now.''} (P17). Similarly, as S94 explained: \emph{``We also get false expectations
of security from such domains. Somebody can generate another onion key with same
facebookcorewwwi address. It's hard but may be possible. People who believe in uniqueness
of generated characters, will be caught and impersonated.''}. Among our survey respondents, there was also concern that the short and recognizable
prefixes tempt users to verify only the prefix and ignore the non-vanity part of
the onion domain, as epitomized by one survey respondent: \emph{``I only
memorize the first part of the domain.''} (S96) while another
wrote: \emph{``If there isn't some cognizable word at the start, it'll be more
difficult for me to determine if I'm going to the correct domain or a scam. I
may end up going to less onion sites as a result.''}~(S355)

This viewpoint was echoed by our interview participants,
who noticed that vanity domains can negatively
affect security. P13 explained: \emph{``I think in theory, on the one [hand], it
makes it easier for you to recognize where you are, it makes it easier for you
to perhaps, share the URL or type it out.  On the other hand, I've seen
concerns that, by having a vanity URL where perhaps people only look
for the Facebook portion and they don't pay attention to what comes after it
could potentially make it easier to exploit unsuspecting users. Send them a link
that also says Facebook but the numbers after it are different, but you just see
the Facebook part and go, `It's fine, it's Facebook.' That can be a risk to
them.''}  P5 also shared their view on vanity domains: \emph{``It seems like it would
encourage more trust on behalf of the user, but then again, maybe make phishing
easier too, if phishers are making vanity domains themselves.  Yeah, that seems
like it could go both ways actually.'' }

\subsubsection{Onion sites are hard to verify as authentic}

\begin{figure}[t]
    \centering
    \includegraphics[width=\linewidth]{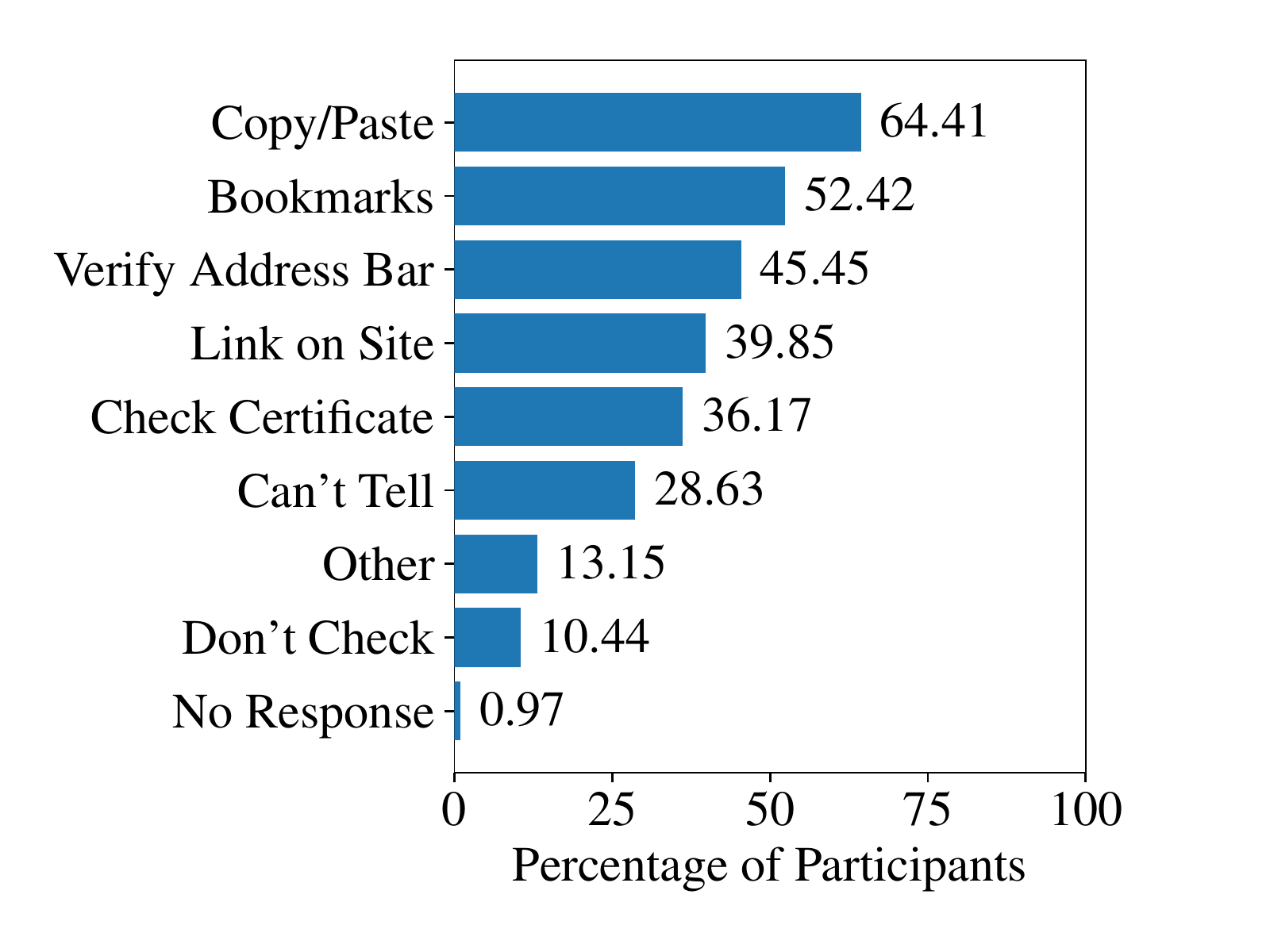}
    \caption{Determining an onion service's legitimacy.}
    \label{fig:determining-legitimacy}
\end{figure}

\paragraph{Verification techniques.} We asked our participants about verifying the authenticity of
an onion site. The majority of our survey respondents (79\%) did want to verify
an onion service as authentic.  \Cref{fig:determining-legitimacy} gives an
overview of the strategies that our respondents employ.  Most of the respondents (64\%) copied and pasted onion links from trusted sources (\eg, friends or another, trusted website) or used
bookmarks when revisiting onion services (52\%). Many survey respondents also verified the
domain in the browser's address bar (45\%), checked if the corresponding website
had a link to its onion site (40\%), or checked that the onion service has a valid
HTTPS certificate (36\%).\footnote{DigiCert is issuing EV
certificates for onion sites~\cite{DigiCert2015a}, but adoption has been
slow---presumably in part because EV certificates require the
CA to verify the applicant's identity and they are not free.} Survey respondents
reporting checking the corresponding regular website for verification, verifying
if familiar images were recognized, or checking for HTTPS (9/517). 8/517 only used
links if received form a trusted resource or trusted member of a community or check
with their notes (4/517).  5/517 trusted their perception of a website as verification
of authenticity or Tor or the fact that onion sites are self-certified by design (3/517) or use the fact that they could log into a site as verification (5/517). Only a few mentioned using multiple sources to verify authenticity (3/517) and at least 9 survey respondents said that they did not use onion links at all.

When asked how many characters our survey respondents verify in onion domains,  19\%
verified thirteen to sixteen digits, \ie, (almost) the full domain, while 20\% 
verified up to nine digits, which is within the realm of brute force attacks, and 5\% verified between nine to twelve digits. More than half of respondents provided no response at all (54\%). 

For those interviewees (7/17) who did attempt to ensure they were visiting an authentic onion site, we observed two strategies: 
relying on someone else to ensure a link was authentic and  trying to work out authenticity using various techniques on their own. Most interviewees in the first group stated that they rely on word of mouth for
verification~(5/17), followed by assistance from someone else~(4/17).  P3 explained
\emph{``[I] let people show me them.  I don't go there myself.''}  Two interview
participants relied on resources they already trusted for onion links, like friends and other communities and two accessed onion services by first visiting their corresponding publicly available websites if they could to verify authenticity.
One of the most common approaches in the second group (3/17) was
to check and compare URLs to see whether they matched to a \emph{``clearnet
site''} (P14), its unencrypted version on the regular Internet.  Furthermore, two interview participants rely on their own
experience, one on HTTPS certificates, and another one would lower the
security settings in Tor Browser using the security slider to check the website
more thoroughly:\emph{``Sometimes, it worries me, but before that I access, in Tor,
I turn off, I always. First, I always turn off the Java service and etcetera, to
check the website. I think it's good, then I will lower the security level in Tor
browser, but mostly, I will ask anything, maybe, in the Reddit or in the forum---in my country forum---of what the service [may be].''} (P17).
One interviewee believed that just using Tor is verification in itself and
another participant avoided onion sites altogether.

\paragraph{Verification challenges.} Indicative of potential security issues, 29\% of survey respondents stated that they sometimes could
not tell the difference between an authentic service and an impersonation, and
10\%  never checked a service's legitimacy in the first place.  Survey
participants who selected ``Other'' (13\%) provided a wide variety of ad-hoc verification
strategies, further highlighting the importance of being able to verify a
site as being the one that they were trying to reach. For instance, 13 survey respondents said there is no good way of verifying onion services or they do not know how to. 

We also asked our interview participants how they knew that the site they went
to was the one that they wanted to visit.  Similar to the survey respondents, six interviewees reported that they did not know how to verify the authenticity on an onion site and they were concerned about being on an impersonating website because it is easy to mistype onion domains and onion domains change frequently if an onion service is short-lived or moves.  P1 summarized the issue as being inherent to the nature of onion services \emph{``I wouldn't know how to do that, no. Isn't that the whole point of onion services? That people can run anonymous things without being to find out who owns and operates them?''} Two
interviewees even believed onion site authentication to be impossible.  For this reason, some interviewees also proposed that onion domain formats without numbers or with a stable patterns of letters and numbers could potentially make sites easier to reach and verify for authenticity.

\subsubsection{Onion lookups suggest typos or phishing}

Phishing remains an issue despite onion services' extra anonymity and security
properties.  Past work has documented phishing onion sites that transparently
rewrote Bitcoin addresses to hijack Bitcoin
transactions~\cite{Winter2016a,Nurmi2015a,Monteiro2016a}.  Key to this attack is
the difficulty of telling apart an authentic onion domain from an impersonation.
For conventional domains we rely on EV certificates, browser
protections, search results, and long-lived reputation, but none of these
methods have matured for onion services.  Does the nature of onion services
facilitate phishing attacks?  If so, what can we do to mitigate the issue?

Most interview participants (9/17) agreed that phishing constitutes a serious
risk, one of them explained the phenomenon this way: \emph{``the two approaches I know
from the normal Web still apply here, which is typo-squatting, registering an
onion [domain] that's only a slight variation away, or bit-squatting, which is
slightly different, but it involves a single or a few bit flips within an onion
address, so that it looks relatively similar''} (P6), while another interview
participant presented their solution to this problem: \emph{``If you're manually
typing it in I suppose they could be a problem, but I primarily cut and paste''}~(P16). 

We evaluated how often lookups to two different onion
domains are extremely similar to one another, which can
shed light on how often an onion domain may be phished, since it is unlikely for
distinct onion services to have extremely similar
strings for onion domains.

To do so, we computed the Jaro-Winkler
similarity metric between each unique pair of correctly formatted onion domains,
which is the edit distance between two
strings that gives more weight to strings with common prefixes. 
We used
this metric because people tend to check the first part of the domain. 
Values range between $[0,1]$, where 0 represents completely different
strings and 1 represents matching strings, to each unique domain pair.  We find that 0.007\% (8,672) of all unique domain
pairs (119,668,185) have an extremely high similarity ($> .90$); for example,
{\tt bitfog2jzic5tnh7.onion} and {\tt bitfog2y7y2pfv75.onion} have a Jaro-Winkler
similarity of 0.917.

We first analyzed the results of the similarity metric for any well-known vanity domains.  We 
found that Facebook's onion site ({\tt facebookcorewwwi.onion}) has a similarity
score of 0.953 with another onion domain that was looked up
{\tt facebookizqekmhz.onion}, which only appeared in our dataset twice (in comparison
to the 101 instances of {\tt facebookcorewwwi.onion}).  
Another frequently looked up onion domain is {\tt blockchainbdgpzk.onion}, which
is a popular Bitcoin wallet; 
it was extremely  similar to blockchatvqztbll.onion (similarity score 0.949).  These
cases of similar domains could be a potential indicator of phishing sites for popular
domains.

\input{jw-table}

We next explored the top 20 most frequently requested onion domains
dataset by
checking: whether they are extremely similar to another onion domain in our dataset, and 
whether there is a large difference in frequency of the two similar domains.  Of the top 20 
onion domains, 16 had a Jaro-Winkler similarity score $>0.90$ with at least 
one other onion domain in the data.  Table \ref{tab:similarity} shows the characteristics of these domains. Many of the domains in the table under ``Onion 1'' are associated with either 
the WannaCry Ransomware, the Mischa Ransomware, or the Petya Ransomware.  The remaining domains in that column are real onion 
domains that returned search results when used as input to \url{https://ahmia.fi};
these include a Russian Market ({\tt hydraruzxpnew4af.onion}), 
DuckDuckGo ({\tt 3g2upl4pq6kufc4m.onion}), and The Hidden Wiki ({\tt zqktlwi4fecvo6ri.onion}).

\subsection{Areas for Improvement}
\label{sec:improve}

When we asked about areas for improvement in the survey and interviews, participants told us that onion services could be enhanced technically and performance-wise, and that privacy and security, educational resources on, and methods for discovering onion content could be improved.

\paragraph{Technical Improvements.} In our open ended question on improvements to
onion services, 43/517 did not provide an answer and 36/517 expressed their gratitude for Tor and Torproject and were satisfied with the service overall. However, many respondents spoke of possible enhancements. The majority of survey respondents (59/517) mentioned technical improvements they would like to see for onion services such as improving support for Javascript, making onion services available in other browsers, and having more support for mobile devices. 17/517 wanted a better user interface and user experience with onion services in general. Our interviewees also mentioned various
technical improvements they would like to see in onion services. Two wanted a secure bookmarking tool and
another interviewee wanted CAPTCHAs to be gone (these are triggered more often with onion services). Only four talked about wanting to see influential websites or even all websites set up corresponding onion sites.

\paragraph{Performance Concerns.} At least 48 survey respondents had performance
concerns about onion services.  For example, one survey user stated, \emph{``I would always prefer the onion site
but for video sites like YouTube I would likely often use the normal site to be
able to get a higher quality stream due to higher bandwidth.''} (S435) Three interview
participants similarly
raised the ``slowness'' of onion services.
 
\paragraph{Privacy and Security.} 34 survey participants expressed concern about
anonymity and security issues and would like to feel and be safer over the Tor network more generally. For instance, S70 wrote: \emph{`I hear a lot of social media questions from casual or unsophisticated users, and the single biggest problem is that they don't have the slightest idea of exactly what's being protected and what isn't. Vague pronouncements that "doing X is safer" don't help. Tor needs to stop being muddy in explaining what it protects, and stop promoting itself to people who don't understand what it can and can't do for them.'} 11/517 complained about lack of anonymity protection specifically from government, big companies or even Federal Bureau of Investigation (FBI). 8/517 wanted to verify onion services as legitimate or live and only 2/517 spoke about not wanting the dark net to contain criminal content.

\paragraph{Education and Resources.} 24 survey respondents believed that there was
a ```knowledge'' issue with not enough resources and documentation for newcomers to Tor and onion services. Many of our interviewees felt similarly
(7/17).  Interviewees lamented about a lack of documentation or resources that would
allow newcomers to learn more about onion services.  P8, for example, wanted to
know how to use onion services correctly and stop being uncertain about its
properties: \emph{``Really clear user education in the installation process would
be great for people like me \ldots who are like `Okay, this is a thing I can
use, why am I using it again? What am I using it for? What does it do?''}
Three of our interviewees also referred to the lack of proper education as \emph{``cultural
mysticism.''}  Uneducated users often misunderstand concepts, as P10 explained: \emph{
``The perception that these are hardcore security tools sometimes signals to
ordinary users that they are also difficult or badly designed or complicated to
use, and that's not really the case with Tor.''}  Even if knowledge was not an
issue, fear of consequences may deter users otherwise, as P8 mentioned before: \emph{``Because it's also
super scary. You think you're playing with this spy thing \ldots Sometimes
it's actually a really simple technical thing that's not terrifying.  And to
 demystify those things would be really nice.''}

\paragraph{Improved Search.} 15/517 survey respondents wanted onion services to
be more accessible, such as via a good search engine or organized database. At least four interviewees also desired improved search engines. As an example of this sentiment, S116 wrote: \emph{`Ask someone to develop a really good search engine so that sites may be found. I am sure that the dark net has to be more than a few illicit sites that are selling stolen credit cards, and running Bitcoin scams. I feel like when I browse the dark net, I am floating in space waiting for another planet to suddenly appear. Whatever content is out there needs to be discovered, lest people will make misinformed judgments about the dark net. The dark net should be understood to be preeminently about privacy, not criminality.'} In addition, many survey
respondents expressed frustration about the difficulty of finding out if a
particular public website has a corresponding onion service.
A common wish was to have a website list its onion service prominently in a
footer or on the corresponding Internet site (3/517).  Ironically, some survey respondents
were surprised that {\tt torproject.org}
has a corresponding onion site---they could not find it on the website.

%% file: jw-table.tex
\begin{table}[t]
	\centering
    \begin{footnotesize}
	\begin{tabular}{l r | l r | r }
	\toprule
	Onion 1 & \# & 
	Onion 2 & \# & 
	J-W \\
	\midrule
	\texttt{57g7spgrzlojinas} & 1,621 & \texttt{57g7spgrziojinas} & 14 & 0.989 \\
	\texttt{xxlvbrloxvriy2c5} & 1,593 & \texttt{xxlvbrioxvriy2c5} & 4 & 0.949 \\
	\texttt{gx7ekbenv2riucmf} & 1,476 & \texttt{gm7ekbenv2riucmf} & 4 & 0.973 \\
	\texttt{mischapuk6hyrn72} & 1,062 & \texttt{mischa5xyir2mrhd} & 8 & 0.902 \\
	\texttt{petya3jxfp2f7g3i} & 1,061 & \texttt{petya3jxfb2f7g3i} & 8 & 0.997 \\
	\texttt{petya3jxfp2f7g3i} & 1,061 & \texttt{petya37h5tbhyvki} & 58 & 0.907 \\
	\texttt{mischa5xyix2mrhd} &  786 & \texttt{mischa5xyir2mrhd} & 8 & 0.999 \\
	\texttt{hydraruzxpnew4af} &  529 & \texttt{hydraruzxpnew1af} & 2 & 0.999 \\
	\texttt{hydraruzxpnew4af} &  529 & \texttt{hydraruehfq5poj5} & 2 & 0.927 \\
	\texttt{hydraruzxpnew4af} &  529 & \texttt{hydraruzxpnew3af} & 2 & 0.999 \\
	\texttt{3g2upl4pq6kufc4m} &  472 & \texttt{tg2upl4pq6kufc4m} & 2 & 0.971 \\
	\texttt{3g2upl4pq6kufc4m} &  472 & \texttt{3g2upl4t5houfo4y} & 2 & 0.924 \\
	\texttt{3g2upl4pq6kufc4m} &  472 & \texttt{3g2upl4oq6kuc4mm} & 2 & 0.954 \\
	\texttt{3g2upl4pq6kufc4m} &  472 & \texttt{3g2upl4pe3kcf24d} & 2 & 0.973 \\
	\texttt{zqktlwi4fecvo6ri} &  410 & \texttt{zqktlwipcfe3siu2} & 2 & 0.931 \\
	\texttt{zqktlwi4fecvo6ri} &  410 & \texttt{zqktlwi4i34kbat3} & 12 & 0.946 \\
	\bottomrule
	\end{tabular}
    \end{footnotesize}
        \caption{The Jaro-Winkler similarity score for frequently visited onion domains in the DNS root dataset.}
    \label{tab:similarity}
\end{table}

%% file: discussion.tex
\section{Future Directions}
\label{sec:discussion}

Our work highlights several opportunities for
improvements to current onion services. 

\paragraph{Security indicators for onion services.} First, many of
our participants had an incomplete mental model of how onion services work and
trusted them less than other Tor services, which suggests that a better
indicator of the protections an onion service offers should be made visible to
onion service users. Currently, The Tor Project is working on a security
indicator for onion services~\cite{trac23247}.  \Cref{fig:onion-service}
illustrates that Tor Browser currently, in version 7.0.10, displays an onion
service connection as an insecure HTTP connection, thus greatly
``under-selling'' the security and privacy that an onion service connection
provides. The design process for such indicators should evaluate whether users
understand the meaning of the indicator, as well as how it differs from an
HTTPS indicator. (Felt \ea found the subtleties that one must consider when
designing similar security indicators~\cite{Felt2016a}.)

\begin{figure}[t]
    \centering
    \includegraphics[width=\linewidth]{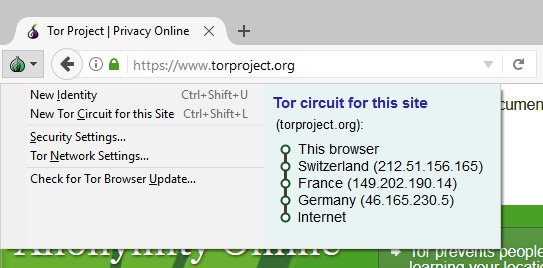}
    \caption{A click on the onion icon reveals the Tor relays that constitute
    the circuit that was used to fetch the current page.  As of February 2018,
    the user interface is subject to a redesign~\cite{trac24309}.}
    \label{fig:tor-button}
\end{figure}

The Tor Browser's circuit display interface is also being redesigned (see
\Cref{fig:tor-button})~\cite{trac24309}.   As with an onion service indicator,
an evaluation of the circuit display could reveal user misunderstandings that
may improve perceptions of and trust in onion services. For example, we found that
some users are not familiar with the concept of guard relays and incorrectly expect
each relay in their circuit to change, which suggests the need for an improved interface.
Users also found it difficult to verify the authenticity of an
onion site; while certificates do help, many sites still do not have
them, and some may never have them.

\paragraph{Automatic detection of phishing onion domains.} Our findings that some
onion domains in the root DNS data have small edit distance to popular onion domains
suggests that users may fall victim typos to phishing attacks; on the other hand,
because the number of popular onion domains is still relatively small and (through
our analysis and previous work~\cite{mohaisen2017leakage,thomas2014measuring}) relatively
well-known, the Tor Browser could raise
an alert when the user attempts to access an onion domain that has a small edit
distance to a popular onion domain.

\paragraph{Opt-in publishing of onion sites.}  Our participants often wanted
more services to be available as onion services and did not often know if an
onion service for a popular website existed.   Participants found it difficult
to discover new onion services, which suggests the need for better ways to
find active onion services. While search engines and curated lists do exist,
they do not generally allow users to locate an onion service of interest
without also stumbling upon unwanted content. One possibility is an opt-in
public log, whereby users can learn about new onion domains as they are added.
Many participants also expressed interest in a browser feature that could
automatically ``upgrade'' from a regular web site to its corresponding onion
service. (The Tor Project is currently investigating this problem
space~\cite{trac21952}.)

\paragraph{Privacy-preserving onion bookmarking.}
Participants found it difficult to track and save onion links; they often
resorted to memorizing links to avoid security issues with storing onion
links. This problem suggests the need for 
a privacy-preserving bookmarking tool that allows users to bookmark sites
without leaving a trail in their browser storage or elsewhere on their system.

%Future work could implement our recommended design suggestions above and evaluate these improvements for onion service users. Future studies could also expand on our work to address the limitations discussed in \ref{sec:limitations} for instance, by recruiting a wider and more varied sample of onion service users where possible.

%% file: conclusion.tex
\section{Conclusion}
\label{sec:conclusion}

Onion services resemble the 1990s web: Pages load slowly, user
interfaces are clumsy, and search engines are inadequate.  Users
appreciate the extra security, privacy, and NAT punching properties of onion
services, which gives rise to a variety of use cases. Yet, users are
confronted with a variety of privacy, security and usability concerns that
should be addressed in future generations of onion services. For example,
users are concerned about the susceptibility of onion domains to phishing
attacks, and the onion domains that are leaked to the public Internet
illustrate that this threat is real---and unaddressed. Users have limited ways
of discovering the existence of onion services, let alone navigating to them.

A range of design improvements, from better discovery mechanisms to automatic
``upgrading'' to a corresponding onion service when it is available are
initial steps to improve usability. Some of these desired features have clear
analogs in the public Internet, such as the padlock icon as a security
indicator for HTTPS, and HTTP Strict Transport Security (HSTS) to automatically
upgrade an HTTP connection
to HTTPS. We expect that many of the usability design lessons 
from the public Internet may in some cases also apply to onion
services.

%Some users found
%ways to manage what is perhaps the most striking idiosyncrasy of onion
%services---their domain format---by using bookmarks, using (encrypted) files
%containing links, and referring to trusted link aggregators.  Many users however
%have flawed mental models of onion services that could increase their
%susceptibility to phishing attacks.

%In this work we studied how Tor users interact with onion services and the
%broader technology that enables onion services.  Drawing on a mixed-methods
%approach, we conducted seventeen semi-structured interviews and collected 828
%responses to our online survey.  These two data sets served as the basis of our
%analysis and provided unique insight into how Tor users perceive, use, and
%understand the Tor network in general and onion services in particular.

%% file: acknowledgments.tex
\section*{Acknowledgments}

This research was supported by the National Science Foundation Awards
CNS-1540066, CNS-1602399, and CNS-1664786.  We thank George Kadianakis for
helpful feedback on our survey questions, Katherine Haenschen for helping us
improve our method, Mark Martinez for conducting 
interviews, Stephanie Whited for helping us disseminate our survey, and
Antonela Debiasi for informing us about current user experience efforts around
the Tor Browser.  We thank Roya Ensafi, Will Scott, Jens Kubiziel, and Vasilis
Ververis for pre-testing our survey, and USC's Information Sciences Institute
for access to the DNS B root data.  We also thank the Tor community
for helpful feedback, for volunteering for our interviews, and for taking our
survey.